\PassOptionsToPackage{svgnames}{xcolor}
\documentclass[sigconf]{acmart}

\setcopyright{acmlicensed}
\copyrightyear{2018}
\usepackage[font=small,skip=4pt]{caption}

\def\overleafhome{/tmp}

\usepackage{subcaption}
\usepackage{fontawesome}

\usepackage{import}
\ifx\homepath\overleafhome
  
\usepackage{etoolbox}

\newtoggle{AlgsNoEnd}

\providebool{tr}
\providebool{conf}

\usepackage[utf8]{inputenc}
\newcommand{\code}[1]{\texttt{#1}}
\usepackage[binary-units=true]{siunitx}
\usepackage{xspace}
\usepackage[svgnames]{xcolor}

\usepackage{soul}
\soulregister\cite7
\soulregister\autoref7
\soulregister\code7
\soulregister\toolname7
\soulregister\ref7
\soulregister\pageref7
\usepackage{hhline}
\definecolor{lightyellow}{RGB}{250, 250, 180}
\definecolor{HLYELLOW}{RGB}{240, 127, 0}
\definecolor{hlyellow}{RGB}{240, 127, 0}
\sethlcolor{lightyellow}

\usepackage{algorithm}
\iftoggle{AlgsNoEnd} {
  \usepackage[noend]{algpseudocode}
}{
  \usepackage{algpseudocode}
}
\usepackage{pifont}
\usepackage{amsmath}
\algdef{SE}[FOREACHIN]{ForEachIn}{EndForEachIn}[2]{\algorithmicfor\ \textbf{each}\ #1\ \textbf{in}\ #2\ \textbf{do}}{\algorithmicend\ \algorithmicfor}%
\algdef{SE}[FOREACHP]{ForEachP}{EndForEachP}[1]{\algorithmicfor\ \textbf{each}\ #1\ \textbf{in parallel do}}{\algorithmicend\ \algorithmicfor}%
\algdef{SE}[FOREACHINP]{ForEachInP}{EndForEachInP}[2]{\algorithmicfor\ \textbf{each}\ #1\ \textbf{in}\ #2\ \textbf{in parallel do}}{\algorithmicend\ \algorithmicfor}%
\algdef{SE}[FORMOD]{ForMod}{EndForMod}[3]{\algorithmicfor\ #1\ \textbf{from}\ #2\ \textbf{to}\ #3\ \textbf{do}}{\algorithmicend\ \algorithmicfor}%
\algdef{SE}[FORMOD]{ForModS}{EndForModS}[4]{\algorithmicfor\ #1\ \textbf{from}\ #2\ \textbf{to}\ #3\ \textbf{step}\ #4\ \textbf{do}}{\algorithmicend\ \algorithmicfor}%
\algdef{SE}[FORMODP]{ForModPS}{EndForModPS}[3]{\algorithmicfor\ #1\ \textbf{from}\ #2\ \textbf{to}\ #3\ \textbf{in parallel do}}{\algorithmicend\ \algorithmicfor}%
\algdef{SE}[FORMODP]{ForModP}{EndForModP}[4]{\algorithmicfor\ #1\ \textbf{from}\ #2\ \textbf{to}\ #3\ \textbf{step} #4\ \textbf{in parallel do}}{\algorithmicend\ \algorithmicfor}%

\iftoggle{AlgsNoEnd} {
  \algtext*{EndFunction}% Remove "end while" text
  \algtext*{EndIf}% Remove "end if" text
  \algtext*{EndFor}% Remove "end if" text
  \algtext*{EndForEachIn}% Remove "end if" text
  \algtext*{EndForEachP}% Remove "end if" text
  \algtext*{EndForEachInP}% Remove "end if" text
  \algtext*{EndForModP}% Remove "end if" text
  \algtext*{EndForModPS}% Remove "end if" text
  \algtext*{EndForModS}% Remove "end if" text
}{}

\algdef{SE}[VARIABLES]{Variables}{EndVariables}{\algorithmicvariables}
  {\algorithmicend\ \algorithmicvariables}
\algnewcommand{\algorithmicvariables}{\textbf{global}}
\algnewcommand{\LineComment}[1]{\State \(\triangleright\) #1}
\algnewcommand{\And}{\textbf{and}\xspace}

\usepackage{tikz}
\usepackage{pifont}
\usetikzlibrary{shapes}

\usepackage{float}
\usepackage{listings}
\newfloat{codeblock}{H}{myc}
\definecolor{darkblue}{rgb}{0,0,.6}
\definecolor{darkred}{rgb}{.6,0,0}
\definecolor{darkgreen}{rgb}{0,.5,0}
\definecolor{red}{rgb}{.98,0,0}
\definecolor{gray}{rgb}{.6,.6,.6}
\definecolor{newgreen}{RGB}{169,209,142}
\definecolor{newpurple}{RGB}{237,134,254}
\definecolor{neworange}{RGB}{244,177,131}
\definecolor{newyellow}{RGB}{255,217,102}
  \usepackage[outputdir=../]{minted}
\else

  \usepackage[frozencache,outputdir=.]{minted}
\fi

\usepackage[utf8]{inputenc}

\usepackage{multirow}
\usepackage{makecell}
\usepackage{enumitem}

\usepackage{pifont}
\newcommand{\checkmarkOurs}{\ding{52}}
\newcommand{\checkmarkNegativeOurs}{\ding{56}}
\usepackage{adjustbox}

\copyrightyear{2025}
\acmYear{2025}
\setcopyright{cc}
\setcctype{by}
\acmConference[SC '25]{The International Conference for High Performance Computing, Networking, Storage and Analysis}{November 16--21, 2025}{St Louis, MO, USA}
\acmBooktitle{The International Conference for High Performance Computing, Networking, Storage and Analysis (SC '25), November 16--21, 2025, St Louis, MO, USA}
\acmDOI{10.1145/3712285.3759868}
\acmISBN{979-8-4007-1466-5/2025/11}

\author[M. Copik]{Marcin Copik}
\affiliation{%
	\institution{ETH Zürich}
	\city{Zürich}
	\country{Switzerland}
}
\email{mcopik@gmail.com}

\author[E. Alnuaimi]{Eiman Alnuaimi}
\orcid{}
\affiliation{%
	\institution{ETH Zürich}
	\city{Zürich}
	\country{Switzerland}
}
\email{ealnuaimi@student.ethz.ch}

\author[A. Kamatar]{Alok Kamatar}
\orcid{}
\affiliation{%
	\institution{University of Chicago}
	\city{Chicago}
	\country{USA}
}
\email{alokvk2@uchicago.edu}

\author[V. Hayot-Sasson]{Valerie Hayot-Sasson}
\orcid{}
\affiliation{%
	\institution{University of Chicago}
	\city{Chicago}
	\country{USA}
}
\email{valerie.hayot-sasson@etsmtl.ca}

\author[A. Madonna]{Alberto Madonna}
\orcid{}
\affiliation{%
	\institution{ETH Zürich}
	\city{Lugano}
	\country{Switzerland}
}
\affiliation{%
	\institution{Swiss National Supercomputing Centre (CSCS)}
	\city{Lugano}
	\country{Switzerland}
}
\email{alberto.madonna@cscs.ch}

\author[T. Gamblin]{Todd Gamblin}
\orcid{}
\affiliation{%
	\institution{Lawrence Livermore National Laboratory (LLNL)}
	\city{Livermore}
	\country{USA}
}
\email{tgamblin@llnl.gov}

\author[K. Chard]{Kyle Chard}
\orcid{}
\affiliation{%
	\institution{University of Chicago}
	\city{Chicago}
	\country{USA}
}
\affiliation{%
	\institution{Argonne National Laboratory (ANL)}
	\city{Chicago}
	\country{USA}
}
\email{chard@uchicago.edu}

\author[I. Foster]{Ian Foster}
\orcid{}
\affiliation{%
	\institution{University of Chicago}
	\city{Chicago}
	\country{USA}
}
\affiliation{%
	\institution{Argonne National Laboratory (ANL)}
	\city{Chicago}
	\country{USA}
}
\email{foster@uchicago.edu}

\author[T. Hoefler]{Torsten Hoefler}
\orcid{}
\affiliation{
	\institution{ETH Zürich}
	\city{Zürich}
	\country{Switzerland}
}
\affiliation{%
	\institution{Swiss National Supercomputing Centre (CSCS)}
	\city{Zürich}
	\country{Switzerland}
}
\email{htor@inf.ethz.ch}



\newtheoremstyle{test}% name
{1.5pt}% Space above
{1.5pt}% Space below
{}% Body font
{}% Indent amount
{\itshape}% Theorem head font
{:}% Punctuation after theorem head
{.5em}% Space after theorem head
{}% Theorem head spec (can be left empty, meaning ‘normal’)

\theoremstyle{test} 

\newtheorem{hyp}{Hypothesis}

\newtheorem{definition}{Definition}

\newcommand{\toolname}{XaaS\xspace{}}

\definecolor{delim}{RGB}{20,105,176}
\definecolor{numb}{RGB}{106, 109, 32}
\definecolor{string}{rgb}{0.64,0.08,0.08}

\lstdefinelanguage{json}{
    numbers=left,
    numberstyle=\small,
    frame=single,
    rulecolor=\color{black},
    showspaces=false,
    showtabs=false,
    breaklines=true,
    postbreak=\raisebox{0ex}[0ex][0ex]{\ensuremath{\color{gray}\hookrightarrow\space}},
    breakatwhitespace=true,
    basicstyle=\ttfamily\small,
    upquote=true,
    morestring=[b]",
    stringstyle=\color{string},
    literate=
     *{0}{{{\color{numb}0}}}{1}
      {1}{{{\color{numb}1}}}{1}
      {2}{{{\color{numb}2}}}{1}
      {3}{{{\color{numb}3}}}{1}
      {4}{{{\color{numb}4}}}{1}
      {5}{{{\color{numb}5}}}{1}
      {6}{{{\color{numb}6}}}{1}
      {7}{{{\color{numb}7}}}{1}
      {8}{{{\color{numb}8}}}{1}
      {9}{{{\color{numb}9}}}{1}
      {\{}{{{\color{delim}{\{}}}}{1}
      {\}}{{{\color{delim}{\}}}}}{1}
      {[}{{{\color{delim}{[}}}}{1}
      {]}{{{\color{delim}{]}}}}{1},
}

\begin{document}

\title{XaaS Containers: Performance-Portable Representation With Source and IR Containers}

\begin{abstract}
	High-performance computing (HPC) systems and cloud data centers are converging, and containers are becoming the default method of portable software deployment.
	Yet, while containers simplify software management, they face significant performance challenges in HPC environments as they must sacrifice hardware-specific optimizations to achieve portability.
	Although HPC containers can use runtime hooks to access optimized MPI libraries and GPU devices, they are limited by application binary interface (ABI) compatibility and cannot overcome the effects of early-stage compilation decisions.
	Acceleration as a Service (XaaS) proposes a vision of \emph{performance-portable} containers, where a containerized application should achieve peak performance across all HPC systems.
	We present a practical realization of this vision through Source and Intermediate Representation (IR) containers, where we delay performance-critical decisions until the target system specification is known.
	We analyze specialization mechanisms in HPC software and propose a new LLM-assisted method for automatic discovery of specializations.
	By examining the compilation pipeline, we develop a methodology to build containers optimized for target architectures at deployment time.
	Our prototype demonstrates that new XaaS containers combine the convenience of containerization with the performance benefits of system-specialized builds.
\end{abstract}

\maketitle

\vspace{-0.75em}
{\small\noindent\textbf{XaaS Implementation:} \url{https://github.com/spcl/xaas-containers}}

{\small\noindent\textbf{XaaS Artifact}: \url{https://doi.org/10.5281/zenodo.17115960}}

\section{Introduction}

High-performance computing (HPC) systems and cloud data centers have been developed to address different goals---HPC aimed at peak performance, while cloud platforms focused on usability.
Recent years have brought a shift towards the convergence of HPC and cloud systems: the architecture of supercomputers is becoming more commoditized~\cite{doi:10.1177/10943420231166608,9810039},
and the popularity of HPC workloads is growing, fueled by the massive demand for machine learning (ML) training.
Cloud introduces new user types and software development philosophies such as containers.
Containers are built on the fundamental assumption that an application is shipped with all its dependencies configured and compiled, regardless of the type of system on which it will be deployed.
Thus, containers offer workload portability, enabling execution at different cloud providers with minimal deployment complexity.
For example, they can efficiently support rapid deployment to spot virtual machines, which are offered at a discount but might not provide the exact hardware specification expected by the user.
Thus, users can decrease computation costs by scaling up their applications on a mix of virtual machines with different hardware configurations that cannot always be predicted.
In HPC, containers can help with seamless deployment across systems with heterogeneous hardware configurations~\cite{7103433}.
However, the adoption of containers in HPC is limited by the lack of \textbf{performance portability}, i.e., the ability to
\emph{"achieve excellent performance on a variety of architectures"}~\cite{hoefler2024xaas}.
\begin{figure}[t]
	\centering
	\includegraphics[width=\linewidth]{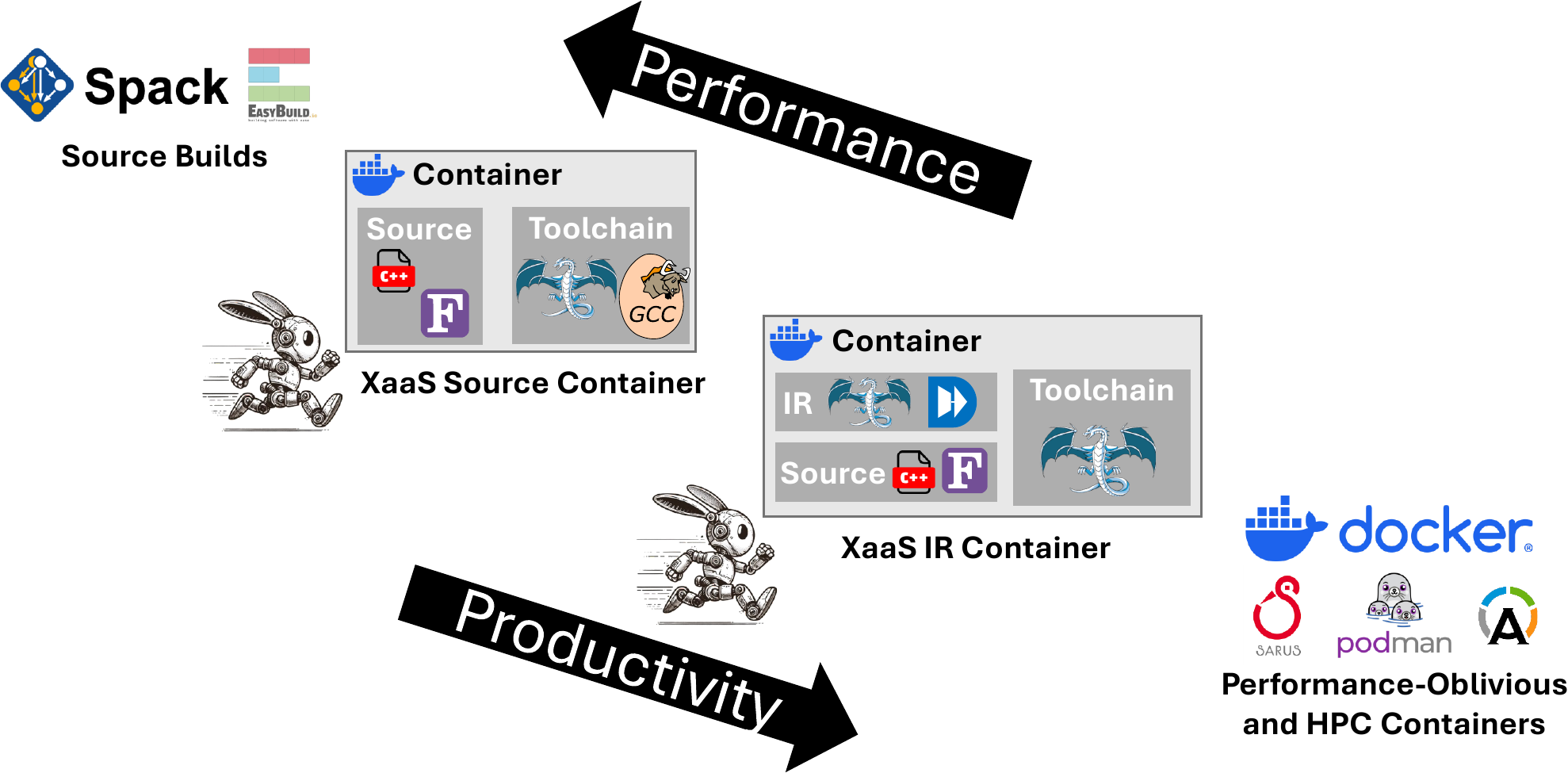}
	\caption{Continuum of software deployment in HPC: performance-portable \toolname{} containers provide better productivity than optimized builds, while avoiding limitations of traditional containers.}
	\vspace{-2em}
	\label{fig:xaas_new_posterchild}
\end{figure}

While containers simplify deployment, they do not necessarily benefit HPC system providers.
%
% from \cite{e4s}
% "A crisis point has been reached where supercomputing centers are employing teams of system professionals just to maintain and upgrade software packages, largely because the software dependencies for such applications typically run deep – down to the specific versions of the Kernel, glibc library and runtime communication library. Significant system resources are now being spent to maintain the dependency tree associated with each application."
Traditional HPC systems use modules containing carefully tailored applications and libraries that allow operators to \emph{nudge} users toward performant solutions.
However, this adds a major human cost of managing HPC software with complex dependencies~\cite{e4s}.
However, when users can simply run any container image, neither they nor administrators will optimize the build, leading to poor performance as users bring ready-to-use but unoptimized software.
To encourage performant solutions, we need a different approach to containers in HPC.

The adoption of containers is also limited by the many parameter configurations that need to be supported, with a combinatorial explosion of containers specialized toward different systems, runtimes, compilers, and parallelism approaches~\cite{10.1145/2807591.2807623}.
HPC containers can be (re)specialized at runtime with hooks defined by the Open Container Initiative (OCI) standard.
These hooks replace libraries inside the container with system-specific versions.
A common example is replacing MPI libraries~\cite{10.1007/978-3-030-34356-9_5},
which requires implementations with compatible Application Binary Interface (ABI)~\cite{10.1145/3615318.3615319}.
The Libfabric installation can be replaced to access proprietary and custom network providers,
which accelerates communication without changes to MPI~\cite{10029965}.
However, solutions using Libfabric are not fully portable due to differences in the capabilities of network providers (Section~\ref{sec:hpc-software-portability}).
Furthermore, Fortran applications are known to lack ABI compatibility, which prevents straightforward runtime replacement of libraries such as BLAS or LAPACK.
Finally, it is often too late to overcome the consequences of prior decisions that affect the generated code, like GPU acceleration and vectorization.

%Instead, 
We propose that containers should be agnostic of selected configurations and target platforms, distributing software packages almost ready for installation while deferring performance decisions until the target system is known.
However, this problem is difficult because HPC build systems are sophisticated, often Turing-complete tools that hardcode performance-critical decisions early in the build process, making complete analysis of these systems not only a massive engineering undertaking but potentially intractable.

\begin{figure}[t]
	\centering
	\includegraphics[width=\linewidth]{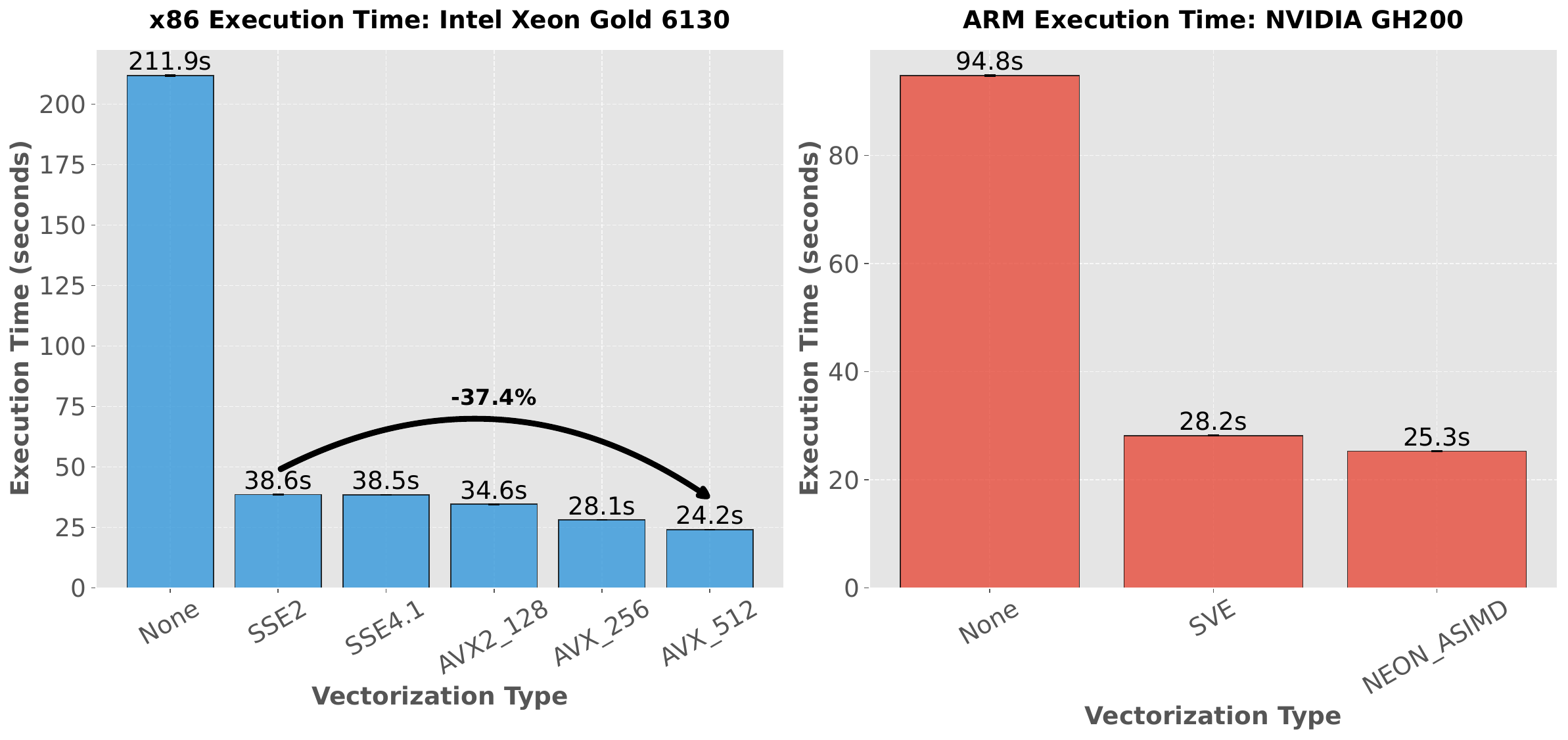}
	\caption{The impact of vectorization in GROMACS (16 threads, 100 timesteps, I/O time excluded): enabling newer features can improve performance, at the cost of creating a non-portable deployment.}
	\vspace{-1.5em}
	\label{fig:gromacs-microbenchmark}
\end{figure}

Acceleration as a Service (XaaS)~\cite{hoefler2024xaas} introduced a new vision of HPC, where \emph{performance-portable} containers can offer the convenience of containers with the performance of specialized builds.
We realize this vision with the concept of \textbf{Source} and \textbf{Intermediate Representation (IR)} containers.
We aim to strike %the middle ground 
a balance between traditional HPC practices, where builds are conducted entirely on the destination system, and limited container optimizations at runtime (Figure~\ref{fig:xaas_new_posterchild}).
We propose to deploy applications in a portable manner and provide the benefits of traditional containers: smaller size, faster deployment, and the ability to hide the application's source code.
Both representations enable configuration for a specific system, allowing us to improve performance by tuning parameters such as vectorization, which enables hardware features unavailable on all platforms (Figure~\ref{fig:gromacs-microbenchmark}).
Instead of distributing \emph{multi-arch} Docker containers targeting different Instruction Set Architectures (ISAs), we distribute \emph{multi-arch-IR} containers to support different compilers and toolchains, as long as they offer an intermediate representation to the end user.

We first analyze the broad world of HPC software to determine their specialization mechanisms (Section~\ref{sec:hpc-software}).
Based on those results, we examine the compilation pipeline and analyze at which levels performance-critical decisions are made and how they can be delayed until the final hardware specification is known (Section~\ref{sec:xaas-containers}).
We configure many instances of the same project with different parameters.
We isolate a shared core of IR files where compilation is identical across configurations or is unaffected by the change in performance-critical parameters (Section~\ref{sec:ir-container}).
We build a container image that is fully optimized and lowered to the target architecture only during \emph{deployment} on the selected HPC system (Section~\ref{sec:ir-integration}).
We demonstrate a prototype of IR containers with LLVM IR~\cite{1281665}.

\newcommand{\PreserveBackslash}[1]{\let\temp=\\#1\let\\=\temp}
\newcolumntype{C}[1]{>{\PreserveBackslash\centering}p{#1}}

\begin{table*}[t!]
\caption{Most important specialization points of selected HPC applications and benchmarks: analysis of representative applications shows a wide diversity of specializations for accelerated and high-performance computing.}
\label{tab:focus-points}
\adjustbox{max width=\textwidth}{
  \begin{tabular}{lcccC{3.3cm}cc}
\toprule
%\textbf{Domain} & \textbf{Name} & \makecell{\textbf{Architecture}\\\textbf{Specializations}} & \textbf{Accelerators} & \textbf{Parallel} & \textbf{Vectorization} & \textbf{Performance Libraries} \\
\textbf{Domain} & \textbf{Name} & \makecell{\textbf{Architecture Spec.}} & \textbf{GPU Acceleration} & \textbf{Parallelism} & \textbf{Vectorization} & \textbf{Performance Libraries} \\
\midrule

Molecular Dynamics & 
GROMACS~\cite{https://doi.org/10.1002/jcc.20291} & 
Architecture-specific FFT & 
OpenCL, CUDA, SYCL, HIP & 
OpenMP, MPI & 
Automatic, many ISAs & 
BLAS/LAPACK, FFT (many) \\

\midrule

Hydrodynamics & 
LULESH~\cite{LULESH:spec} & 
- & 
- & 
OpenMP, MPI & 
- & 
- \\

\midrule

Electronic Structure & 
\makecell{Quantum\\Espresso~\cite{Giannozzi_2009}} & 
Compiler adaptations & 
CUDA, OpenACC & 
OpenMP, MPI & 
- & 
\makecell{BLAS/LAPACK, ELPA~\cite{AUCKENTHALER2011783},\\ScaLAPACK~\cite{234898}, FFT (many)} \\

%& & & & & &\\

\midrule

\multirow{3}{*}{Lattice QCD} & 
MILC~\cite{doi:10.1177/109434209100500406} & 
Compiler adaptations & 
CUDA, HIP, SYCL & 
OpenMP, MPI & 
\makecell{Compiler flags, many ISAs\\(Intel, AMD, PowerPC)} &
\makecell{LAPACK, PRIMME~\cite{10.1145/1731022.1731031},\\FFTW~\cite{1386650}, QUDA~\cite{CLARK20101517,quda}}\\

%\midrule
\cmidrule[0.5pt]{2-7}

%Lattice QCD & 
& 
OpenQCD~\cite{openqcd} & 
Optimized for x86 CPUs & 
- & 
OpenMP, MPI  & 
Assembly (SSE, AVX, FMA3) & 
- \\

\midrule

Particle-in-Cell & 
\makecell{VPIC~\cite{10.1063/1.2840133},\\VPIC 2.0~\cite{9444146}} & 
Kokkos portability & 
CUDA & 
%\makecell{OpenMP, pthreads,\\MPI} & 
OpenMP, MPI &
\makecell{OpenMP and V4 library\\(many ISAs)} & 
- \\

\midrule

Cloud Physics & 
CloudSC~\cite{cloudsc} & 
System-specific toolchains & 
CUDA, SYCL, HIP, OpenACC & 
OpenMP, MPI & 
- & 
Atlas~\cite{DECONINCK2017188} \\

\midrule

Weather \& Climate & 
ICON~\cite{icon1} & 
System-specific toolchains & 
CUDA, HIP, OpenACC & 
OpenMP, MPI & 
System-specific compiler flags &
BLAS/LAPACK \\

\midrule

LLM Inference & 
llama.cpp~\cite{llamacpp} & 
Optimization flags & 
\makecell{Eight, including CUDA,\\HIP, SYCL} & 
OpenMP, pthreads & 
\makecell{Intrinsics (AVX, AVX2,\\AVX512, AMX, NEON, \dots)} & 
BLAS (OpenBLAS, MKL, BLIS~\cite{BLIS1}) \\

\bottomrule
\end{tabular}}
\vspace{-1em}
\end{table*}

\begin{table*}[t!]
	\caption{Levels of code portability and their implementations. Libraries like MPI can be shipped with a container and replaced dynamically at runtime or fully mounted inside the container during execution, as long as ABI compatibility is provided.}
	\label{tab:portability-layers}
	\adjustbox{max width=\textwidth}{
	%\begin{tabular}{p{2cm}p{2.5cm}p{3cm}p{3cm}p{3cm}}
	\begin{tabular}{lcccc}
		\textbf{Level}                                                                          & \textbf{Technology} & \textbf{Description} & \textbf{Portability Approach} & \textbf{Dependency Integration} \\
		\midrule

		Building                                                                                &
		Spack~\cite{10.1145/2807591.2807623}, EasyBuild~\cite{6495863}                          &
		From-source package manager                                                             &
		Parameterized package compilation                                                       &
		Automatic, dependency resolver                                                                                                                                                                         \\

		\midrule

		Linking                                                                                 &
		Sarus~\cite{10.1007/978-3-030-34356-9_5}, Apptainer~\cite{10.1371/journal.pone.0177459} &
		HPC container runtime                                                                   &
		Runtime binding, OCI hooks                                                              &
		Manual, CLI option, and host bind                                                                                                                                                                      \\

		\midrule

		Lowering                                                                                &
		Linux Popcorn~\cite{10.1145/3093337.3037738}                                            &
		Multi-ISA binary system                                                                 &
		Heterogeneous-OS containers                                                             &
		No direct integration                                                                                                                                                                                  \\

		                                                                                        &
		H-containers~\cite{10.1145/3381052.3381321,10.1145/3524452}                             &
		ISA-agnostic container with IRs                                                         &
		Container + recompilation                                                               &
		No direct integration                                                                                                                                                                                  \\

		                                                                                        &
		NVIDIA PTX                                                                              &
		Runtime JIT compilation                                                                 &
		Virtual GPU architecture                                                                &
		No direct integration                                                                                                                                                                                  \\

		\midrule

		Emulation                                                                               &
		Wi4MPI~\cite{9556086}, mpixlate~\cite{mpixlatecray}                                     &
		MPI compatibility layer                                                                 &
		Runtime emulation of MPI ABIs                                                           &
		No direct integration                                                                                                                                                                                  \\
	\end{tabular}}
\end{table*}

In this paper, we make the following contributions:
\begin{itemize}
	\item We study specialization points in HPC applications and propose reorienting container deployment around them.
	\item We analyze the multi-layer HPC compilation stack, and design two solutions for performance-portable containers.
	\item We introduce IR containers that combine the convenience of generic containers with performance of specialized builds.
\end{itemize}

\section{State of HPC Software}
\label{sec:hpc-software}

To understand the challenge of performance portability in HPC, we begin with identifying configuration parameters that affect the performance by \emph{specializing} to the target machine (Section~\ref{sec:hpc-software-specialization}).
Then, we analyze the existing approaches for code portability, focusing on \emph{when} the optimizations are applied (Section~\ref{sec:hpc-software-portability}).
This process helps us decide which build steps should be conducted before distributing software to the end user (Section~\ref{sec:xaas-containers}).

\subsection{Specialization Points}
\label{sec:hpc-software-specialization}

HPC applications are highly configurable since they aim to run on heterogeneous systems, with many built on custom and specialized hardware (Table~\ref{tab:focus-points}).
We define \textbf{specialization points} as \emph{application parameters that must be known at the configuration and build time, stay constant throughout the entire application's lifetime, and whose values affect the final performance and portability}.
In particular, we focus on parameters that dictate which specific hardware and software solutions should be employed by the final application.
These options are not always mutually exclusive, as the application can support multiple GPU backends that are only selected at runtime.
We consider the following categories of specialization points.
\begin{itemize}[topsep=0pt]
	\item Network fabric and communication library like MPI.
	\item Acceleration, such as NVIDIA, AMD, or Intel GPUs.
	\item CPU-specific optimizations such as vectorization.
	\item Libraries like BLAS, LAPACK, and FFT. % have different implementations with varying performance on various platform.
\end{itemize}

\vspace{-1em}

\subsection{Portability Layers}
\label{sec:hpc-software-portability}

%We categorize existing approaches to HPC portability solutions into four broad categories based on the portion of the application build that is conducted on the target system (Table~\ref{tab:portability-layers}).
We classify portability solutions into four categories based on the fraction of the build that is conducted on the target system (Table~\ref{tab:portability-layers}).

\textbf{Building} performs a full compilation of the application on the destination system.
This approach provides the highest performance portability, at the cost of increased complexity - each user builds their copy manually or with the help of a package manager.

In \textbf{linking}, the dynamic dependencies of an existing application are replaced at runtime with an optimized and systems-specific implementation, e.g., through OCI hooks for containers.
The main constraint here is the requirement of ABI compatibility, which prevents such replacements for BLAS/LAPACK libraries.
For example, Libfabric allows the implementation of network communication with a standardized API and dynamic selection of network providers at runtime~\cite{githublibfabric}.
In practice, it still requires manual and specialized implementations because network providers differ in the support of libfabric features (Table~\ref{tab:libfabric-features}).
Furthermore, while libfabric replacement can accelerate a containerized MPI runtime~\cite{10029965}, it might require additional plugins to support intra-node messaging~\cite{https://doi.org/10.1002/cpe.8203}.
Thus, relinking the libfabric installation is not a general method for performance specialization of an already compiled application.

\begin{figure*}[t!]
	\centering
	\begin{minipage}[t]{0.24\textwidth}
		\centering
		\begin{minted}[fontsize=\footnotesize]{cpp}
#if not defined(HAVE_ANY_BLAS)
void transpose(double* A,
 double* B, int rows, int cols) {
 for (int i; i < rows; i++) {
  for (int j; j < cols; j++) {
   B[j * rows + i] = 
    A[i * cols + j];
  }}
}
#endif
        \end{minted}
		\textbf{(a)} Manual Implementation
	\end{minipage}
	\hfill
	\begin{minipage}[t]{0.24\textwidth}
		\centering
		\begin{minted}[fontsize=\footnotesize]{cpp}
#if defined(HAVE_OPENBLAS)
void transpose(double* A,
 double* B, int rows, int cols) {
 cblas_domatcopy(
  CblasRowMajor, CblasTrans,
  rows, cols, 1.0, A,
  cols, B, rows
 );
}
#endif
        \end{minted}
		\textbf{(b)} OpenBLAS Implementation
	\end{minipage}
	\hfill
	\begin{minipage}[t]{0.22\textwidth}
		\centering
		\begin{minted}[fontsize=\footnotesize]{cpp}
#if defined(HAVE_MKL)
void transpose(double* A,
 double* B, int rows,
 int cols) {
 mkl_domatcopy(
  'R', 'T', rows, cols,
  1.0, A, cols, B, rows
 );
}
#endif
        \end{minted}
		\textbf{(c)} Intel MKL Implementation
	\end{minipage}
	\hfill
	\begin{minipage}[t]{0.28\textwidth}
		\centering
		\begin{minted}[fontsize=\footnotesize]{cpp}
#if defined(HAVE_CUBLAS)
void transpose(double* A, double* B,
    int rows, int cols) {
 double alpha = 1.0, beta = 0.0;
 cublasDgeam(handle, CUBLAS_OP_T,
  CUBLAS_OP_N, rows, cols, &alpha, A,
  cols, &beta, nullptr, rows, B, rows
 );
}
#endif
        \end{minted}
		\textbf{(d)} cuBLAS Implementation
	\end{minipage}
	\caption{Matrix transposition: a simple linear algebra kernel that has not been standardized, requiring a custom solution chosen at build time. While manual implementation is a safe choice that will work everywhere, it will prevent achieving the highest performance.}
	\vspace{-1em}
	\label{fig:transpose-implementations}
\end{figure*}

\textbf{Lowering} replaces the intermediate representation with the final binary product at the target system.
These solutions support multiple ISAs, even when the hardware popularity changes over time.
HPC applications cannot be limited to x86 deployments, with primary examples of contenders being PowerPC in the past and ARM today, e.g., Fugaku's A64FX~\cite{9355239}, Graviton CPUs~\cite{9835543} in AWS cloud, and NVIDIA's Grace Hopper superchip~\cite{10.1145/3636480.3637097}.
Similarly, this approach provides compatibility with different NVIDIA GPU architectures by deploying Parallel Thread Execution (PTX), an ISA for virtual GPU architectures in CUDA.
PTX is JIT-compiled to a binary code, providing portability across many GPU generations~\cite{cudadocsvirtualarch}.

Finally, \textbf{emulation} attempts to patch incompatibilities at runtime without code modifications.
An example is replacing MPI runtimes when the application has been built against MPI that is not ABI compatible with the host implementation~\cite{9556086,mpixlatecray,schnetter_2022_6174409}.

\begin{table}[t]
	\footnotesize
	%\caption{Comparison of feature availability in libfabric 2.0~\cite{githublibfabric} providers (\textbf{P} - partial support, \textbf{N/A} - not used, \textbf{?} - unknown). While the library offers a portable communication API, practical implementations must still specialize for available hardware.}
	\caption{Feature availability in libfabric 2.0~\cite{githublibfabric} providers (\textbf{P} - partial support, \textbf{N/A} - not used, \textbf{?} - unknown). Libfabric offers a portable API, but implementations must still specialize to the hardware.}
	\label{tab:libfabric-features}
	%\adjustbox{max width=\linewidth}{
	\centering
	\begin{tabular}{l|c|c|c|c|c}
		\textbf{Feature}    & \textbf{TCP}           & \textbf{IB}            & \textbf{Slingshot}     & \textbf{EFA}           & \textbf{Omni-Path}     \\
		                    & \textbf{(tcp)}         & \textbf{(verbs)}       & \textbf{(cxi)}         & \textbf{(efa)}         & \textbf{(opx)}         \\
		%\multicolumn{6}{l}{\textbf{Endpoint}} \\
		\hline
		Message             & \checkmarkOurs         & \checkmarkOurs         & \checkmarkNegativeOurs & \checkmarkNegativeOurs & \checkmarkNegativeOurs \\
		Reliable Datagram   & \checkmarkOurs         & \textbf{P}             & \checkmarkOurs         & \checkmarkOurs         & \checkmarkOurs         \\
		Datagram            & \checkmarkNegativeOurs & \checkmarkOurs         & \checkmarkNegativeOurs & \textbf{P}             & \checkmarkNegativeOurs \\
		%\hline
		%\multicolumn{6}{|l|}{\textbf{Data Transfer}} \\
		\hline
		% FI_TAGGED
		Tagged Message      & \checkmarkOurs         & \textbf{P}             & \checkmarkOurs         & \checkmarkOurs         & \checkmarkOurs         \\
		% FI_DIRECTED_RECV
		Directed Receive    & \checkmarkOurs         & \checkmarkNegativeOurs & \checkmarkOurs         & \checkmarkOurs         & \checkmarkOurs         \\
		% FI_MULTI_RECV
		Multi Receive       & \checkmarkOurs         & \checkmarkNegativeOurs & \checkmarkOurs         & \checkmarkOurs         & \checkmarkOurs         \\
		% FI_ATOMIC
		Atomic Operations   & \checkmarkNegativeOurs & \textbf{P}             & \checkmarkOurs         & \textbf{P}             & \checkmarkOurs         \\
		%                     &                        &                        &                                     &                        &                        \\
		Memory Registration & \textbf{N/A}           & Basic                  & Scalable               & Local                  & Scalable               \\
		\hline
		%
		%\textbf{Progress \& Events} & & & & &\\
		% FI_PROGRESS_MANUAL
		Manual Progress     & \checkmarkNegativeOurs & \checkmarkNegativeOurs & \checkmarkOurs         & \checkmarkOurs         & \checkmarkOurs         \\
		% FI_PROGRESS_AUTO
		Auto Progress       & \checkmarkOurs         & \checkmarkOurs         & \checkmarkNegativeOurs & \checkmarkNegativeOurs & \textbf{P}             \\
		Wait Objects        & \checkmarkOurs         & \textbf{P}             & \checkmarkOurs         & \checkmarkNegativeOurs & \textbf{?}             \\
		% FI_RMA_EVENT
		Completion Events   & \checkmarkOurs         & \checkmarkNegativeOurs & \checkmarkOurs         & \checkmarkNegativeOurs & \checkmarkNegativeOurs \\
		\hline
		%\textbf{Scalability Features} & & & & &\\
		% FI_RM_ENABLED
		Resource Management & \checkmarkOurs         & \textbf{P}             & \checkmarkOurs         & \textbf{P}             & \checkmarkOurs         \\
		% SCALABLE ENDPOINTS
		Scalable Endpoints  & \checkmarkNegativeOurs & \checkmarkNegativeOurs & \checkmarkNegativeOurs & \checkmarkNegativeOurs & \checkmarkOurs         \\
		% FI_TRIGGER
		Trigger Operations  & \checkmarkNegativeOurs & \checkmarkNegativeOurs & \checkmarkOurs         & \checkmarkNegativeOurs & \checkmarkNegativeOurs \\
	\end{tabular}
	\vspace{-1.5em}
\end{table}

\section{HPC Specialization in XaaS}
\label{sec:xaas-containers}

To fundamentally change the way we distribute HPC software, we first need to understand how \emph{specialization points} affect the build and installation process (Section~\ref{sec:containers-build-specialization}).
Since discovering specialization points is complex due to the lack of standardization in build systems, we apply semi-automatic detection with the help of artificial intelligence (Section~\ref{sec:containers-specialization-discovery}).
By detecting specialization points, we can design performance-portable containers that delay the impact of specialization until we know the final specification (Section~\ref{sec:ir-container}).

\begin{figure*}[t!]
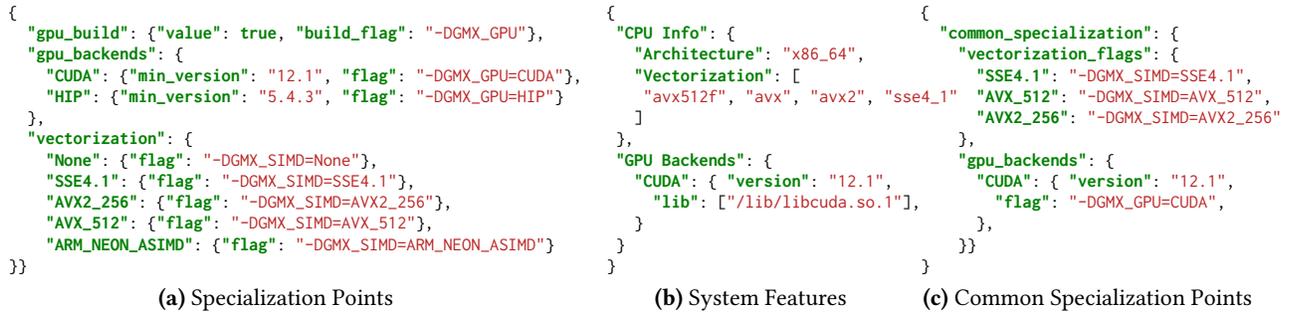

    \centering
    \begin{minipage}[t]{0.4\textwidth}
        \centering
        \begin{minted}[fontsize=\footnotesize]{json}
{
  "gpu_build": {"value": true, "build_flag": "-DGMX_GPU"},
  "gpu_backends": {
    "CUDA": {"min_version": "12.1", "flag": "-DGMX_GPU=CUDA"},
    "HIP": {"min_version": "5.4.3", "flag": "-DGMX_GPU=HIP"}
  },
  "vectorization": {
    "None": {"flag": "-DGMX_SIMD=None"},
    "SSE4.1": {"flag": "-DGMX_SIMD=SSE4.1"},
    "AVX2_256": {"flag": "-DGMX_SIMD=AVX2_256"},
    "AVX_512": {"flag": "-DGMX_SIMD=AVX_512"},
    "ARM_NEON_ASIMD": {"flag": "-DGMX_SIMD=ARM_NEON_ASIMD"}
}}
        \end{minted}
        \textbf{(a)} Specialization Points
    \end{minipage}
    \hspace{0.55cm}
    \begin{minipage}[t]{0.23\textwidth}
        \centering
        \begin{minted}[fontsize=\footnotesize]{json}
 {
  "CPU Info": {
    "Architecture": "x86_64",
    "Vectorization": [
     "avx512f", "avx", "avx2", "sse4_1"
    ]
  },
  "GPU Backends": {
    "CUDA": { "version": "12.1",
      "lib": ["/lib/libcuda.so.1"], 
    }
  }
 }
        \end{minted}
        \textbf{(b)} System Features
    \end{minipage}
    \hspace{0.05cm}
    %\hfill
    \begin{minipage}[t]{0.25\textwidth}
        \centering
        \begin{minted}[fontsize=\footnotesize ]{json}
{
  "common_specialization": {
    "vectorization_flags": {
      "SSE4.1": "-DGMX_SIMD=SSE4.1",
      "AVX_512": "-DGMX_SIMD=AVX_512",
      "AVX2_256": "-DGMX_SIMD=AVX2_256"
    },
    "gpu_backends": {
      "CUDA": { "version": "12.1",
        "flag": "-DGMX_GPU=CUDA",
      },
    }}
}
        \end{minted}
        %\vspace{8.3mm}
        \textbf{(c)} Common Specialization Points
    \end{minipage}
    \caption{Simplified example of GROMACS' specialization points and how the checker determines the intersection of specialization points.}
    \vspace{-1em}
    \label{fig:specialization-points}
\end{figure*}

\subsection{HPC Specialization}
\label{sec:containers-build-specialization}

The build process of an application can be split into three major parts:
\textbf{configuration} that resolves dependencies and decides what should be built, and how;
\textbf{compilation and linking}, responsible for turning source files into libraries and executables;
and \textbf{installation}, which places headers, binaries, and project resources in a selected destination.
To create a transparent and seamless experience for HPC users, any solution must support all three steps.

During \textbf{configuration}, source modules and files are enabled or disabled depending on the selected specialization.
Compiler flags are adjusted, and the build system adds compile-time definitions embedded into the application.
Paths to dependencies are resolved, and additional packages can be fetched into the build directory.

Once source files are \textbf{compiled}, headers of chosen libraries will be introduced, preprocessing directives are applied, and compile-time definitions like C++ templates are resolved.
After that stage, we can no longer switch between libraries that are not ABI-compatible since the application has been introduced to types with different representations and functions with incompatible signatures.
Furthermore, preprocessing directives can potentially exclude certain code paths and already decide which kernels will be generated, as shown in the example of matrix transpose in BLAS libraries (Figure~\ref{fig:transpose-implementations}).
Since this operation is not standardized, different implementations are needed, but they can only be enabled if the selected library is present in the system.

Once the source files are translated into the intermediate representation and optimized, the ISA is chosen, processor-specific decisions are made, and the final code is emitted.
At this point, the code is no longer portable between different systems.
Furthermore, it is no longer feasible to change vectorization settings or apply optimizations
valid only on specific types of CPUs.

At the \textbf{linking} stage, applications are relinked to a specific implementation of a dependency.
This decision can be changed later, as long as the library is linked dynamically and its implementations are ABI compatible.
For example, an application compiled against MPICH can be relinked to use Cray's specialized MPICH implementation.
While future MPI implementations will be ABI compatible~\cite{10.1145/3615318.3615319}, this method is currently limited since MPI types can have different implementations.
After that point, the only possible performance adjustments are runtime options, such as switching network providers in applications built on top of libfabric.

Finally, the application is \textbf{installed}, which includes copying the contents of the package.
Specialization affects the generation of project-specific headers and the installation of libraries, since the inclusion of specific dependencies is affected by user decisions.

\subsection{Specialization Discovery}
\label{sec:containers-specialization-discovery}
To generate the list of specialization points an application supports, we need to parse build scripts and understand what dependencies and optimizations can be selected during configuration.
Unfortunately, this process is not standardized in common HPC programming languages, like C++ and Fortran.
In addition to supporting different build systems such as autotools, handwritten Makefiles, CMake, Bazel, or custom scripts, there is often no single way of determining dependencies within one ecosystem.
For example, third-party libraries can be located in CMake using standard CMake calls such as \texttt{find\_package}, with custom find modules for libraries not supported by CMake, by using \texttt{pkg-config}, or with a manual search for specific headers and libraries.
Moreover, large projects often define custom routines for locating packages.

Analyzing configuration files to identify specialization points is difficult to automate due to the many diverse and unique patterns.
At the same time, it is a task that humans can handle easily. 
Thus, we employ a Large Language Model (LLM) to help users identify specialization points by processing the project configuration files with a structured prompt.
We apply \emph{in-context learning} by including in the prompt examples of specialization options, build flags, and CMake commands, helping the LLM to extract specialization options accurately and capture all relevant choices in the build file.
The model outputs a JSON file containing the detected specialization points. 
To enforce consistency and facilitate automated processing, we supply a predefined JSON schema, guiding the model to adhere to a structured format and minimizing anomalies. % in its responses.
As the accuracy and correctness of LLM systems vary heavily, the results of LLM extraction still serve mainly as a guideline for the developer to prepare the final specification (Section~\ref{sec:eval-llms}).

On the target system, we collect information on system features and available specialization points.
Then, we intersect these results with the specialization discovery of the application.
At this point, we exclude the non-supported configuration options and present the user with a list of options for each specialization point.
Figure~\ref{fig:specialization-points} illustrates a subset of GROMACS’s specialization points alongside the system features of our test environment.
GROMACS supports OpenCL, SYCL, HIP, and CUDA as GPU backends, whereas the system is limited to CUDA and OpenCL.
The automatic checker identifies the intersection of supported GPU backends, and allows the user to manually select the final specialization points.

\section{XaaS Containers}
\label{sec:ir-container}

\begin{figure}[t]
	\centering
	\includegraphics[width=\linewidth]{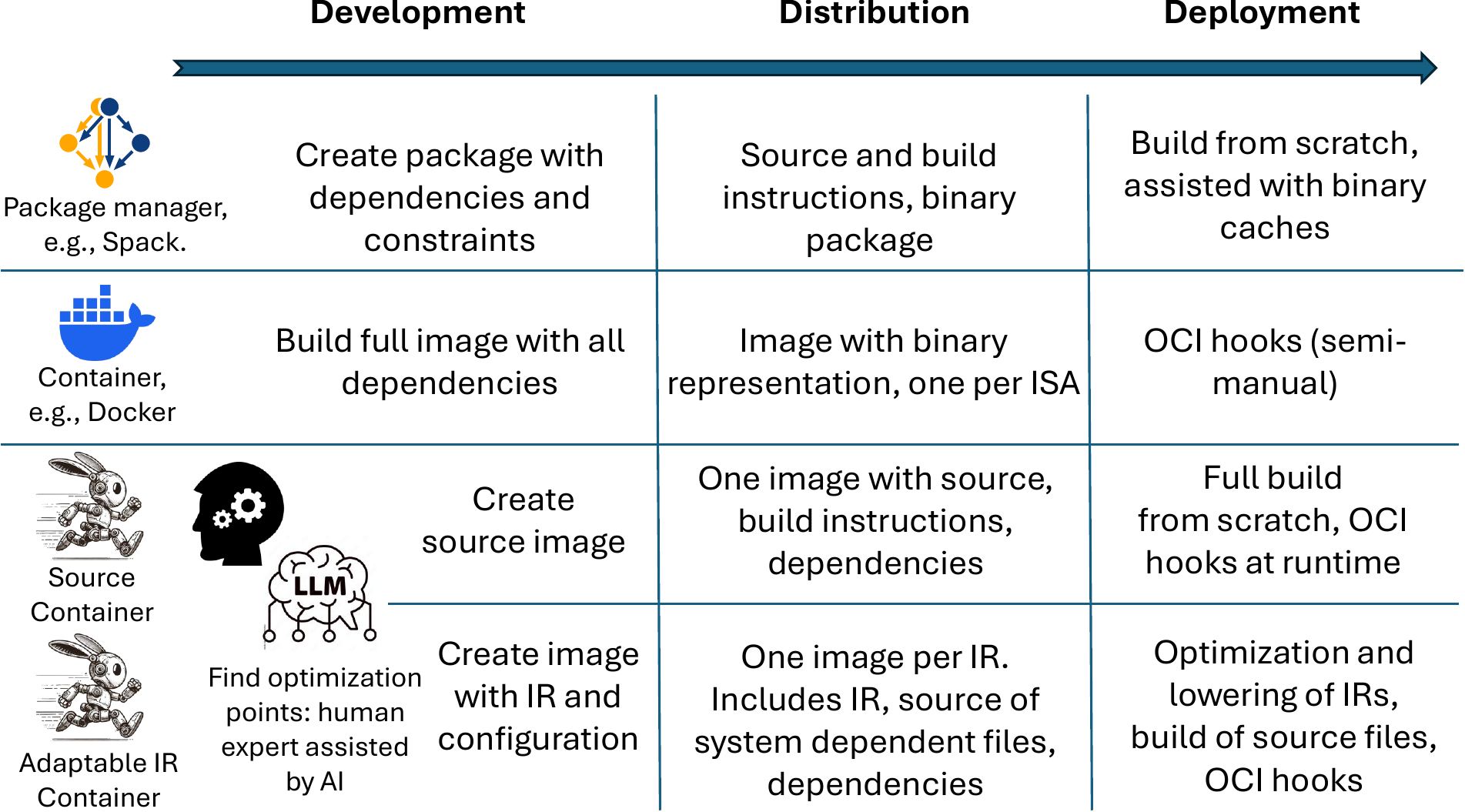}
	\caption{Performance-portable \toolname{} containers provide better productivity while avoiding limitations of traditional containers.}
	\vspace{-1em}
	\label{fig:xaas_spectrum}
\end{figure}

In \toolname{}, we aim to resolve the two major limitations of existing containers---lack of performance portability and a combinatorial explosion of the number of final representations.
First, we deploy \textbf{source containers} that bring the application and its environment to the final system (Section~\ref{sec:xaas-source-containers}).
The source container images contain the HPC application with development tools (Figure~\ref{fig:xaas_spectrum}), and are only built for the target system once hardware configuration and all dependencies are known.
Then, we propose that \emph{intermediate representation} becomes a new mode for distributing software (Section~\ref{sec:xaas-ir-containers}).
Intuitively, we distribute a container image where build steps are conducted until we cannot progress further without making performance-critical decisions (Section~\ref{sec:xaas-ir-containers-pipeline}).

The new types of containers require a new \textbf{deployment step}, when specialization points are matched against the system specification and user preferences.
The remaining source files are compiled, architecture-specific optimizations are applied, and the entire application is lowered to the selected ISA.
As a result, we obtain a new image different from the one provided in the registry, which allows for specialization of the application to the selected HPC system.

\begin{figure}[t]
	\centering
	\includegraphics[width=\linewidth]{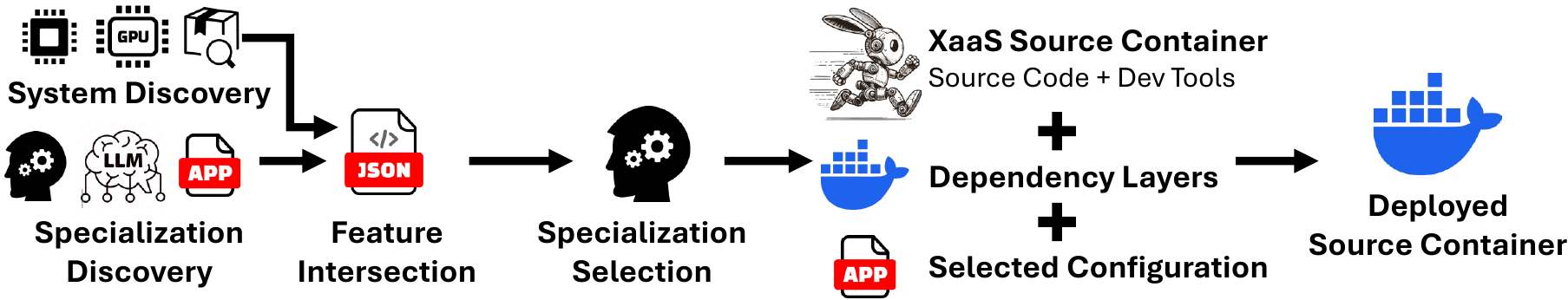}
	\caption{Deployment of source containers on HPC system.}
	\vspace{-1.5em}
	\label{fig:xaas-source-pipeline}
\end{figure}

\subsection{Source Containers}
\label{sec:xaas-source-containers}
Source containers deliver the application source code, an open-source MPI implementation, and the build toolchain to the HPC system.
This solution can support HPC applications and systems that benefit from specialized and vendor-provided compilers, which often do not expose their intermediate representation explicitly.
Since no build steps are conducted before the deployment, this approach does not suffer from the large number of combinations: only one image is needed per toolchain and architecture.

The \textbf{deployment} begins by automatically detecting CPU features, accelerators, and the development environment (Figure~\ref{fig:xaas-source-pipeline}).
This step must be conducted on a compute node, and in an environment with all standard modules loaded.
We augment the results with knowledge of standard HPC environments.
For example, when a ROCm or CUDA installation is discovered, we assume the availability of rocFFT and cuFFT, respectively, even if they are not explicitly detected.
The discovery can be enhanced with solutions for labeling microarchitectural features, e.g., archspec~\cite{9297044}, and strengthened with explicit system specification provided by system operators.

Then, we perform the automatic intersection of specializations (Section~\ref{sec:containers-specialization-discovery}), and the user selects the best fit from the available options.
After that, we generate a Dockerfile to create a new image that inherits from the source container and builds the application with selected options.
We implement support for a subset of popular dependencies, inheriting dependency versions from the system environment when possible, and provide them as Docker layers or build steps.
Other dependencies could be supported by employing package managers like Spack.
Furthermore, we allow switching base images at deployment times to use optimized and recommended images for a specific platform, e.g., oneAPI images in Aurora (Section~\ref{eval:sec-source-containers}).

The new container is no longer portable and can often only be executed on that specific system.
However, images derived from source containers should achieve near-native performance since we enable specializations available for bare metal applications, and the performance losses can only come from the container runtime itself.
By providing the infrastructure for building and storing a single deployed container, we avoid the situation where users manually build multiple copies of the same application.
From the user's point of view, the entire process is still convenient and relatively automatic---only a \emph{cold pull} takes longer than a traditional container build since the very first user of a container on a system will have to wait for the build to finish.
Users are only expected to select the values for discovered specialization points.
However, this step could also be accelerated by allowing system operators to supply preferred configurations, e.g., preferring MKL on Intel systems over other BLAS/FFT libraries, relying on third-party configuration like in Spack~\cite{10.1145/2807591.2807623}, or providing the AI system with the application documentation to suggest the best option for the target platform.

\begin{figure*}[t]
	\centering
	\includegraphics[width=\textwidth]{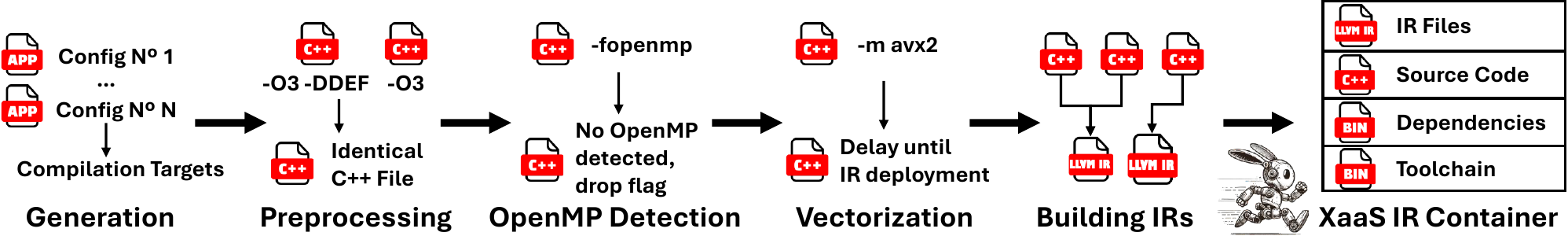}
	\caption{\toolname{} IR container. The modular pipeline reduces the build cost by detecting IR files shared by different configurations.}
	\vspace{-1em}
	\label{fig:xaas-source-pipeline-build}
\end{figure*}

\vspace{-1em}

\subsection{IR Containers}
\label{sec:xaas-ir-containers}
IR containers are close to the original idea of containers, with the main goal of \emph{build once and run anywhere}.
However, the original build is augmented with the deployment step, responsible for the final optimizations and lowering to the target architecture.
The application is distributed in the compiler's intermediate representation, and the image should not contain any object code that depends on the final architecture, as this would be neither portable nor performance-portable.
In addition to selecting the architecture of the container image, we specify the \emph{IR}, e.g., LLVM IR.

IR containers can suffer the same problem of combinatorial explosion that affects performance-oblivious containers.
With multiple possible choices for parallelization, acceleration, hardware specialization, and communication, the number of \emph{build configurations} grows combinatorially.
The cost of building containers that include all combinations would be too high for many applications, and their size would be a major deployment problem.
To make IR deployments practical, we need to deduplicate build configurations and build only the \emph{unique} intermediate representation files:

\begin{hyp}
	Let $P_{1}, P_{2}, \dots, P_{N}$ be $N$ different configurations of the same HPC application.
	Each configuration $P_{i}$ compiles $T_{i}$ different IR files.
	Let $T'$ be the total number of \textbf{distinct} IR files produced in all $N$ configurations.
	Then, $T' < \sum_{i} T_{i}$.
\end{hyp}

The set of unique results of compilation is not immediately detectable, as different compilation settings will obscure the analysis while not affecting the result.
Many build configurations apply compilation flags globally to targets, e.g., C/C++ include flags or enabling OpenMP.
To resolve the problem of too many build configurations, we apply a \emph{behavioral} approach.
Due to the complexity and intractability of the problem, we do not attempt to understand what build systems do but examine the compilation instructions of each target created in build configurations.
We identify the differences between configurations and build only the delta when selecting a specific option.
When two different configurations produce different targets from the same input, we build two different IR files for that target.
Instead of storing all results of many different builds, we use one common set of IR files shared across all configurations, and a set of deltas applied only to selected configuration.

\subsubsection{System Dependency}
Before assembling the IR container, we define the necessary conditions for deploying the application's source files in that form.

\begin{definition}
	A system-independent source file ($SI$) can be passed through the configuration and compilation stages without specifying the final software and hardware configuration.
\end{definition}

A typical HPC example of such a source file would be numerical computations.
Computations can be parallelized with OpenMP since the file can be compiled twice to IR, once with and once without OpenMP.
However, MPI dependencies are not permitted for this category due to the lack of ABI compatibility in current runtimes.
In practice, such files can be compiled with MPICH to be deployed with a widely accepted binary interface~\cite{mpichabi}.
CUDA's PTX can be included in this category, while the binary representation of a compiled kernel (cubin) cannot.

\begin{definition}
	A system-dependent source file ($SD$) cannot be compiled to a shared IR without sacrificing portability.
	%due to inherent dependency on changing interfaces. 
\end{definition}

This category includes files with functionality conditionally enabled only for some configurations,
and files requiring a dedicated compiler that does not expose its intermediate representation.

\begin{hyp}
	HPC applications can be decomposed into two sets of source files: system-independent ($SI$) and system-dependent ($SD$).
	Most importantly, for most practical applications, $|SI| >> |SD|$.
\end{hyp}

The corollary of the last part of the hypothesis is critical: the effort of building a specialized pipeline makes sense only if the majority of the source code can be processed without knowing the final system; otherwise, source containers are a better solution.

\vspace{-0.75em}

\subsection{IR Containers Pipeline}
\label{sec:xaas-ir-containers-pipeline}

We create a modular container build pipeline (Figure~\ref{fig:xaas-source-pipeline-build}) that solves multiple problems to determine the unique set of IR files:
\begin{enumerate}
	\item Combinatorial explosion of build configurations on projects with many specialization points.
	\item Code modules that can be excluded during the project's configuration, depending on specialization points.
	\item C/C++ preprocessor that can encode the effects of specialization points.
	\item Compilation flags that do not affect the result.
\end{enumerate}
In particular, we implement optimizations that analyze the effects of OpenMP and vectorization flags.

\textbf{Configuration}
While we try to constrain the cost of building a container (Problem 1), we need to ensure that we do not prematurely exclude code modules that might become necessary during the deployment step (Problem 2).
First, we generate a specialized build configuration for each combination of provided specialization points, e.g., LULESH~\cite{LULESH:spec} with two specialization points---MPI and OpenMP---will produce four different configurations.
Each build configuration is created in a containerized environment, where the build directory is always mounted under the same path.
This helps remove the effects of different locations on the generated compilation flags.
The container is assembled from layers that provide the toolchain and dependencies of the selected application.

For each build configuration, we obtain the list of all compilation steps and associated compilation flags, e.g., by examining the compile commands database generated by CMake, which can be obtained without analyzing the internal structure of each build system.
Other build systems can be supported by extracting compilation flags with third-party tools and compiler wrappers.
We identify \emph{compilation targets} and not source files, since a single source file can be mapped to multiple targets but with different compilation flags.
Then, we compare the results of each build profile to identify the common denominator, a shared core of files that are always built in the same manner.
In the case of LULESH, where each build consists of five source files, we obtain 20 IR files.

\textbf{Preprocessing}
The configuration step is followed by a preprocessor evaluation to determine if different compile-time definitions produce a semantically different source file.
Thus, we create preprocessed C/C++ files, hash them, and look for identical files.
In LULESH, this step does not change the result since enabling MPI changes the source files, and the OpenMP compilation flag is attached to all files.
Since not all files will use OpenMP, we apply a Clang AST analysis pass to detect if the processed file contains any OpenMP constructs.
If two compilation targets from different build configurations have the same hash but differ only in the OpenMP flag, then we can treat them as the same.
After that step, LULESH has been reduced from 20 to 14 different IR files.

% If we had space, we could have a small SAXPY example:
% clang -O3 -mavx512f produces vector instructions of AVX-512
% clang -O3 -emit-llvm + opt -O3 -mattr=+avx512f + lowering produces vector instructions of AVX (likely wrong vectorized loop structure)
% clang -O3 -mllvm -disable-llvm-optzns + opt -O3 -mattr=+avx512f + lowering produces AVX-512 (identical to first case)
Vectorization is another example of divergence across many builds of the same application.
For example, GROMACS~\cite{https://doi.org/10.1002/jcc.20291} supports nine different configurations for vectorization on x86 CPUs.
Since LLVM vectorizers work at the IR level, we can ignore the architecture-specific flags when comparing different configurations of the same file.
Instead, the vectorization will be applied during deployment once the final ISA is known.
Our experiments show that LLVM optimizations need to be delayed as well, as otherwise the code will not be re-vectorized efficiently once the target is known.

\textbf{Compilation}
During compilation, applications become aware of types that might introduce incompatibilities between different implementations of the same library.
Thus, we need to isolate them from the rest of the codebase: the "default" partition ($SI$) can continue compilation as previously, while the new partition dependent on the ABI-problematic library ($SD$) will not be compiled at all until the final deployment for the target system.

MPI applications are the most important source of ABI compatibility, and we compile against MPICH to provide wide portability.
Since we expect future MPI runtimes to be ABI compatible~\cite{10.1145/3615318.3615319}, we do not focus on this problem.
To support Open MPI, XaaS containers could detect all files dependent on MPI and compile them multiple times against different ABIs, or employ portability layers~\cite{mukautuva}.

\begin{figure}[t]
	\centering
	\includegraphics[width=\linewidth]{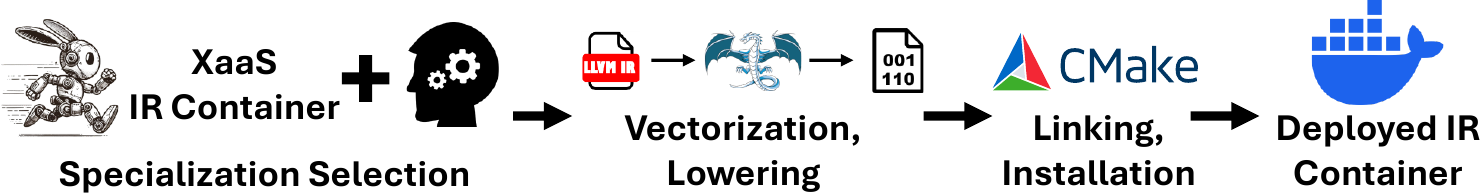}
	\caption{Deploying \toolname{} IR container: user selects one configuration that will be optimized and lowered to the architecture.}
	\vspace{-1.5em}
	\label{fig:xaas-ir-pipeline-deploy}
\end{figure}

\begin{figure}[t]
	\centering
	\includegraphics[width=\linewidth]{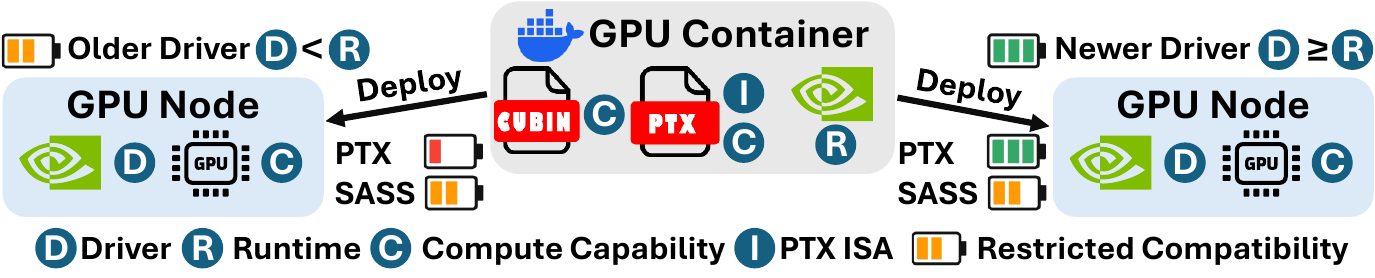}
	\caption{CUDA compatibility is determined by six parameters: two on host (driver and device capability), and four in container (runtime, PTX version, compute capability of PTX and device binary cubin).}
	\vspace{-1em}
	\label{fig:xaas-ir-gpu}
\end{figure}

\textbf{Container Build:}
We generate a container with the LLVM, all build directories, IR files, and the source repository.
The latter is necessary to support system-dependent files and perform the final installation.
For each build configuration, we generate a specific installation file with instructions for compiling IR files and placing them in their respective locations.
We do not include all image layers, e.g., GPU runtimes, as these will be reassembled at deployment.

\textbf{GPU Compatibility:}
GPU runtimes target multiple architectures by generating either direct device code or portable PTX ISA for later JIT compilation (Figure~\ref{fig:xaas-ir-gpu}).
Portable containers can use the oldest supported CUDA runtime to ensure backward compatibility, while newer runtimes offer updated libraries and support for new hardware features at the cost of additional compatibility steps.
We provide compatibility across CUDA minor versions, e.g., CUDA 12.x.
First, we search for any use of compile-time definition indicating CUDA runtime version, which is a pessimistic check if the application might depend conditionally on API features unavailable in older drivers.
Once we decide if the newest CUDA runtime can be used, we emit device binaries for all architectures and a PTX for the latest compute capability to support newer devices.

\subsubsection{Deployment}
The user selects specialization points from the list of parameters and their values chosen at configuration time.
Then, we create a new container by assembling dependencies explicitly defined for that specialization.
We select a subset of IRs for that configuration, optimize and compile them, and let the build system finish linking (Figure~\ref{fig:xaas-ir-pipeline-deploy}).
Image tag includes specialization points to support the coexistence of many builds.

\textbf{Code Generation:}
We lower all IR files of a selected build configuration to the target architecture.
This step can be much faster than a complete compilation of a C/C++ application.
We also apply vectorization at this stage if it is detected during container build.

\textbf{Linking:}
Once the binary code is generated, we can use the existing project configuration to link them together into final libraries and executables.
Alternatively, linking flags for each target can be inferred from the build system, e.g., CMake exposes them explicitly.
For runtime replacement system-optimized libraries, we can rely on the capabilities of existing HPC containers.

\section{XaaS Containers in Practice}
\label{sec:ir-integration}

We implement a prototype of \toolname{} containers that can create source and IR images and then deploy them on selected HPC systems.
For common choices of specialization points, like CUDA or Intel oneAPI, we provide an extensible fleet of Docker containers and manual installation steps.
In source containers, we build a toolset for matching system specifications and specialization points, implement application-specific patching and integration, and provide two base source images, one for x64 and one for ARM64.
The prototype of IR containers is built on top of Clang 19 and CMake, and includes a collection of Docker images with common runtimes and application dependencies.
Users provide application-specific parameters and build steps, from which we generate all build configurations.

The new types of containers proposed in this work differ fundamentally from existing approaches, which raises new challenges in handling different applications and systems (Section~\ref{sec:ir-integration-challenges}).
We transition from multi-arch container images to multi-IR images, highlighting that performance portability requires a change in container structure.
Source and IR containers are vessels for delivering the correct environment and application, and they need to be transformed during the deployment step.
\toolname{} containers will need new approaches to integrate into container ecosystems (Section~\ref{sec:ir-integration-oci}).

\subsection{Challenges}
\label{sec:ir-integration-challenges}

\textbf{Can an IR container be cross-platform?}
\toolname{} needs to create one IR container per architecture, e.g., \emph{IR, x86} and \emph{IR, AArch64}.
While the LLVM intermediate representation can be independent of the target system, this condition does not hold for practical compilation of C/C++ applications.\footnote{C/C++ code \emph{cannot} be compiled to a platform-independent LLVM IR~\cite{llvmdocs}.} 
The IR is affected by the compilation platform, e.g., through type sizes, definitions included in system headers, inlined assembly, and intrinsics~\cite{llvmbitcodeindependent}.

\vspace{0.5ex}
\noindent\textbf{How to handle custom targets?}
Applications can use custom targets to fetch dependencies or generate source files.
For example, when no FFT implementation is selected for GROMACS, it will build a custom implementation, but this does not happen at configuration time - only at build time.
We assume that the user specifies all such targets, and we execute them before analyzing build configurations.

\vspace{0.5ex}
\noindent\textbf{Which IRs are available?}
The IR container requires a toolchain that can export the intermediate representation and import it in subsequent compilation steps.
Here, LLVM IR is the primary example.
While GNU Compilers export the program representation in GIMPLE, this format cannot be imported later and lowered to the target architecture.
On GPUs, the intermediate representation can be provided through PTX on NVIDIA architectures, and SPIR-V for applications using SYCL and OpenCL. 
However, at this moment, the intermediate representations of Intel DPC++/C++ Compiler and Cray Compiling Environment are unavailable to end users.
When partial compilation to IR is impossible, source containers offer the fallback option.
\toolname{} can also use high-level intermediate representations suitable for HPC optimizations, e.g., DaCe SDFG~\cite{10.1145/3295500.3356173}.

\subsection{Compatibility with OCI Containers}
\label{sec:ir-integration-oci}

%\noindent\textbf{OCI Compatibility}

% https://github.com/opencontainers/image-spec/blob/main/spec.md#understanding-the-specification
% https://specs.opencontainers.org/image-spec/
Our deployment model fundamentally differs from traditional containers: \toolname{} completely breaks the relationship between the image in the registry and the image on the system.
This can raise the question of OCI compliance, since the container standard requires that changes to image layers are recorded in the manifest, leading to a new hash value and a new immutable identifier of the image~\cite{ocimagespec}.
However, \toolname{} publishes standard container images, pulls them from container registries, and produces specialized images in the same OCI-compliant format that can be consumed later by general-purpose and HPC-focused container runtimes.
We introduce a new deployment tool customized for HPC specialization, but all other steps of container management - building, publishing, pulling, and running - are conducted with standard and existing container tools.
Furthermore, virtually none of the current HPC container solutions preserve OCI compliance: images are generally flattened~\cite{Gerhardt_2017,10.1007/978-3-030-34356-9_5,10.1145/3126908.3126925} (destroying the original OCI layers), converted to SquashFS, or use the custom Singularity Image Format (SIF)~\cite{10.1371/journal.pone.0177459}.

\vspace{0.5ex}
\noindent\textbf{Image Architecture and Annotations:}
In \toolname{} containers, we propose that the source and IR formats become a new identifying feature of the container image.
This would require that the OCI specification recognizes LLVM IR as a valid architecture.
The current specification allows an image to have an architecture and a variant of the architecture~\cite{ocimagespec}.
%
% https://github.com/opencontainers/image-spec/blob/main/image-index.md#platform-variants
%While variants can be flexible, they must be associated with a specific architecture.
%
Additionally, it reserves a list of \emph{features}
which can be used to encode deployment format.

% https://github.com/opencontainers/image-spec/blob/main/annotations.md
%
OCI images use annotations for additional metadata in various media types (indexes, manifests, image configurations), with the latter consumed directly by container runtimes. 
In \toolname{}, annotations could embed specialization points of the HPC application.
We propose that future versions could include specialization points as image annotations, allowing \toolname{} tools to query them before pulling and building the final image.
Furthermore, it would simplify image tags and allow for the easy location of specialized images. %on the HPC systems.

\section{Evaluation}

\begin{table*}[t!]
  \caption{Performance and cost of LLMs parsing GROMACS configuration. Token counts, latency, and estimated cost are averaged from 10 runs executed from Z{\"u}rich, Switzerland. F1, precision, and recall metrics are aggregated across all runs and reported as Min/Median/Max per model.}
\centering
\label{tab:LLMs-compact}
\renewcommand{\arraystretch}{0.95}
\setlength{\tabcolsep}{4pt}
\begin{adjustbox}{max width=\textwidth}
\begin{tabular}{@{}l l l l l 
                r r r   % F1
                r r r   % Precision
                r r r@{} % Recall
              }
\toprule
%\textbf{Model Version} & \textbf{Avg Token(In)} & \textbf{Avg Token(Out)} & \textbf{Avg Time (s)} & \textbf{Cost (\$)} & 
\textbf{Model Version} & \textbf{Tokens} & \textbf{Tokens Out} & \textbf{Time (s)} & \textbf{Cost (\$)} & 
F1$_\text{min}$ & F1$_\text{med}$ & F1$_\text{max}$ & 
P$_\text{min}$ & P$_\text{med}$ & P$_\text{max}$ & 
R$_\text{min}$ & R$_\text{med}$ & R$_\text{max}$ \\
\midrule

gemini-flash-1.5-exp & 15803 ± 0 & 2333.5 ± 147.6 & 16.40 ± 1.00 & 0.002 & 
0.863 & 0.902 & 0.942 & 
0.838 & 0.898 & 0.948 & 
0.868 & 0.909 & 0.936 \\

gemini-flash-2-exp & 15803 ± 0 & 2610.8 ± 189.4 & 11.96 ± 0.86 & 0.003 & 
0.873 & 0.978 & 0.994 & 
0.873 & 0.981 & 1.000 & 
0.873 & 0.981 & 0.988 \\

claude-3-5-haiku-20241022 & 17841 ± 0 & 1568.9 ± 174.2 & 20.09 ± 1.96 & 0.021 & 
0.628 & 0.672 & 0.702 & 
0.807 & 0.863 & 0.894 & 
0.5 & 0.539 & 0.622 \\

claude-3-5-sonnet-20241022 & 17841 ± 0 & 1528.7 ± 39.2 & 126.18 ± 335.31 & 0.077 & 
0.661 & 0.672 & 0.692 & 
0.875 & 0.878 & 0.882 & 
0.539 & 0.544 & 0.57 \\

claude-3-7-sonnet-20250219 & 17841 ± 0 & 3122.7 ± 155.1 & 50.29 ± 21.67 & 0.100 & 
0.878 & 0.883 & 0.911 & 
0.857 & 0.868 & 0.923 & 
0.9 & 0.9 & 0.9 \\

o3-mini-2025-01-31 & 13538 ± 0 & 8003.9 ± 1160.8 & 108.40 ± 40.02 & 0.050 & 
0.559 & 0.924 & 0.968 & 
0.549 & 0.909 & 0.987 & 
0.57 & 0.923 & 1.000 \\

gpt-4o-2024-08-06 & 13539 ± 0 & 1540.0 ± 146.1 & 26.06 ± 6.96 & 0.049 & 
0.547 & 0.774 & 0.887 & 
0.508 & 0.892 & 0.979 & 
0.53 & 0.675 & 0.859 \\

\bottomrule
\end{tabular}
\end{adjustbox}
\end{table*}

In the evaluation, we focus on the following research questions:
\begin{itemize}[topsep=0pt]
	\item Can LLM systems process large C++ project configurations?
	\item Can source containers provide performance portability?
	\item Can IR containers perform better than portable containers?
	\item Can IR containers optimize deployments?
\end{itemize}

\vspace{-1em}

\subsection{Benchmarking Setup}

We demonstrate the portability of our containers on three systems:
\begin{itemize}[topsep=0pt]
	\item \textbf{CSCS Ault} Heterogenous system with Sarus~\cite{10.1007/978-3-030-34356-9_5} containers. We use Ault23 (Intel 6130 CPU, V100 GPU), Ault25 (AMD EPYC 7742, A100 GPU), and Ault01-04 (Intel 6154 CPU).
	      %with a RoCE network and
	\item \textbf{CSCS Alps.Clariden} Cray system with the GH200 superchip, Cray Slingshot network, and Podman~\cite{10030014} containers.
	\item \textbf{Aurora} Cray system with Intel Xeon CPU Max CPUs, Intel Data Center GPU Max, and Apptainer~\cite{10.1371/journal.pone.0177459} containers.
\end{itemize}

The deployment images for Clariden are built on the compute nodes, and we use a local development machine with Docker for Ault23 and Aurora, as neither system supports container building.

\indent We consider two high-performance applications as case studies.

\noindent\textbf{GROMACS 2025.0}~\cite{https://doi.org/10.1002/jcc.20291}
supports many vectorization options, with automatic detection of the best option on the system.
In addition to MPI and OpenMP, it contains mutually exclusive GPU backends. % for CUDA, SYCL, and OpenCL.

\noindent\textbf{llama.cpp}~\cite{llamacpp}
The C++ LLM inference engine achieves good portability by separating the implementation into multiple backends, which can be loaded dynamically at runtime.

\subsection{Specialization Discovery}
\label{sec:eval-llms}
We propose automating the process of discovering specialization points with the help of LLMs.
Given LLM's tendency to hallucinate and produce incorrect output, this raises the question: can these results be trusted?
Thus, we evaluate the analysis of GROMACS with models from OpenAI, Anthropic, and Google.
For each model, we apply \emph{in-context learning} by providing examples from CMake configuration options, internal commands, and flags for GROMACS, Quantum Espresso, and Kokkos.
We normalize the structure of specialization points, and compare specializations found in the ground truth and LLM result, counting true/false positives and negatives.
We repeat the prompt 10 times and evaluate the characterization of the build system, vectorization, FFT and linear algebra libraries, parallel computing solutions, and GPU backends.
Results in Table~\ref{tab:LLMs-compact} show that the correctness and performance vary widely between models.
Both GPT models produce inconsistent results across repetitions (F-score 0.55--0.97).
Gemini models perfom best, which can be explained by the large context window.
Examples of failures include returning only a subset of options (Claude 3.5, GPT-4o) and mixing FFT and linear algebra libraries (GPT-4o, Gemini 1.5).

% measured with gemini tokenizer
The basic CMake configuration contains 13299 tokens, while all CMake scripts add 154,946 more tokens.
Recommended configurations are described in the documentation that adds almost 1M tokens (without plots), which might be out of reach for many current models~\cite{liu-etal-2024-lost}.
Documentation can be processed iteratively, but this would significantly increase processing time and further increase the uncertainty of the final result.
Thus, while in-context learning can enable LLM models to perform well in the automatic analysis of specialization points,
the processing pipeline requires tight human supervision and corrections, and performance preferences might be
provided by system operators or application developers.

\emph{Generalization}
We evaluated llama.cpp for which we provide no prompt examples.
We pass CMake configurations of llama.cpp and its main subproject \emph{ggml} (2544 and 4574 tokens, respectively).
Best performing models are Sonnet 3.7 (F1 0.55--0.62) and GPT-o3 (F1 0.62--0.7).
Models often underperform due to minor discrepancies (inconsistent hyphen/underscore, missing \emph{-D} prefix).
Normalization improves performance: Sonnet 3.7 (F1 0.63--0.74), Gemini Flash 2 (F1 0.53--0.79), and GPT-o3 (F1 0.73--0.79).
Additionally, ggml contains over 20 optimization flags; including them in evaluation decreases overall performance while improving Sonnet's best result to 0.78.

\begin{figure}[t]
	\centering
	\includegraphics[width=\linewidth]{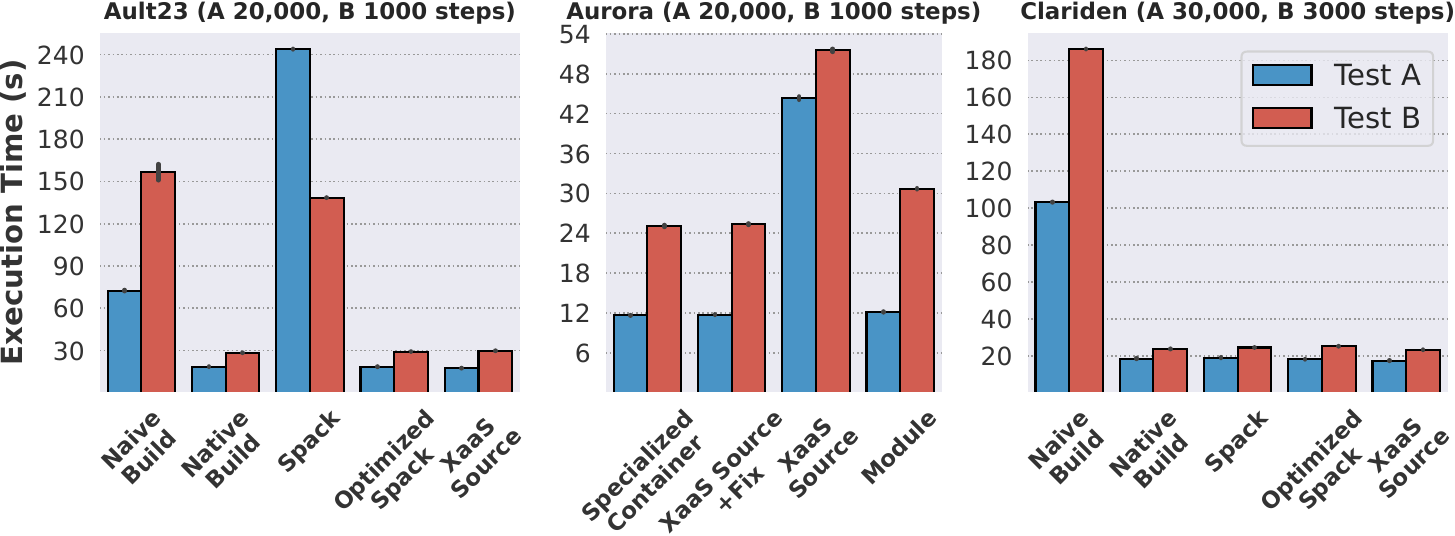}
	\vspace{-1em}
	\caption{Performance portability of GROMACS between systems.}
	\vspace{-1em}
	\label{fig:gromacs-execution time}
\end{figure}

\begin{figure}[t]
	\centering
	\includegraphics[width=\linewidth]{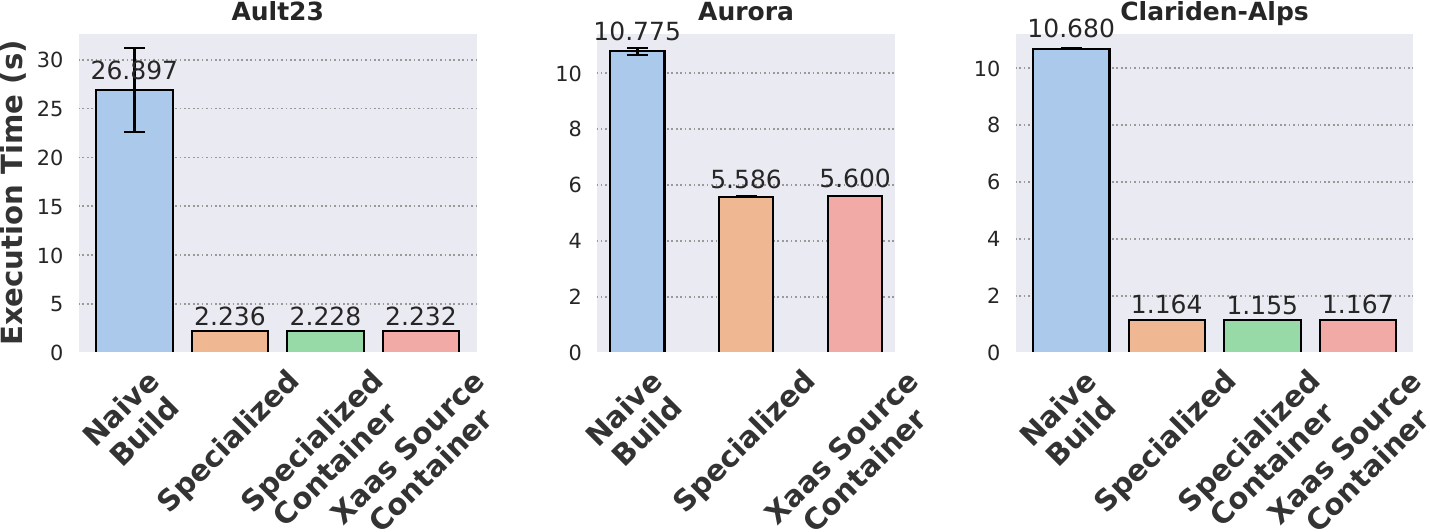}
	\vspace{-1em}
	\caption{Performance portability of llama.cpp between systems.}
	\vspace{-1em}
	\label{fig:llamacpp-execution time}
\end{figure}

\subsection{Performance Portability}
\label{eval:sec-source-containers}

We evaluate the performance portability of \toolname{} source containers by comparing them against local builds, specialized containers, and Spack packages or modules - when available.
We deploy different source images based on system discovery and user input results.

\subsubsection{GROMACS}
We execute test cases A and B from UEABS~\cite{ueabs} 30 times, and subtract the I/O overhead reported by GROMACS from timings (Figure~\ref{fig:gromacs-execution time}).
We use default GCC 11.4 in XaaS source images, and Spack-installed GCC 11.5 for other test cases.
Naive build uses the default CMake command from the documentation, which results with lack of GPU acceleration even when the CUDA module is loaded.
Both naive and native builds pick up MKL from the HPC modules environment.
On Ault23, we compare Spack installation with MPI and CUDA, and a second configuration with explicit selection of MKL, which achieves performance close to the XaaS source container.
According to GROMACS logs, the default Spack installation performs worse in the CPU part of the application, indicating possible issues with multithreading or the automatically selected OpenBLAS.
%
% Processing ault23 - B - gromacs_testcase4_testcaseB, samples: 30, 30
% Processing ault23 - testcase4, time: 138.48863333333335 +- 0.04346232419715079
% Processing ault23 - B - gromacs2025_testcase4_testcaseB, samples: 30, 30
% Processing ault23 - testcase4, time: 136.72633333333334 +- 0.057986244274133014
%
% Processing ault23 - B - gromacs2025_testcase5_testcaseB, samples: 30, 30
% Processing ault23 - testcase5, time: 28.391166666666667 +- 0.04138302141366523
% Processing ault23 - B - gromacs_testcase5_testcaseB, samples: 30, 30
% Processing ault23 - testcase5, time: 29.283299999999997 +- 0.04099056083102681
Spack cases uses the latest available GROMACS 2024.4, and a subsequent reevaluation of 2025.0 with test B on Ault23 demonstrated an average improvement of 1-2 seconds.

We use two baselines on Aurora: a hand-written specialized container and a module version of GROMACS 2024.5 since it cannot be installed by Spack.
For XaaS and specialized containers, we use the oneAPI image recommended by system operators.
The module version uses MPI, while other benchmarks use the internal Threads-MPI due to MPI compatibility issues in containers on Aurora (Section~\ref{sec:network_performance}).
However, the default source container uses only CPUs because the build is incompatible with Intel Max GPUs.
There, GROMACS uses a compile-time definition specialized only for this device, found in the documentation but not the build configuration.
For the \emph{manual} fix, source containers need an additional source of knowledge, such as documentation parsing (Section~\ref{sec:eval-llms}) or specialization parameters provided by developers (Section~\ref{sec:xaas-source-containers}).

\textbf{Portable Container.}
A GPU-capable container is possible only with the SYCL backend.
This \emph{exotic} configuration~\cite{gromacsexotic} uses the vkFFT library~\cite{10036080}, which can be compiled for one hardware backend only.
Instead, we used the recently added oneMath library~\cite{onemath}, which supports MKL and cuFFT simultaneously.
We built GROMACS with the CUDA plugin for SYCL~\cite{codeplaycudaplugin} in the oneAPI image, and compared it against our source containers on V100 (Ault23) and A100 (Ault25).
The SYCL container is 11\%-20\% slower on test A, and fails to run test B.
Portability is further limited by two factors: GPU compatibility still requires compile-time definitions, and the CUDA plugin generates code for one GPU architecture at a time.

\subsubsection{llama.cpp}
We use the internal llama.cpp benchmark, running 40 repetitions of prompt processing (512) and text generation (128), with a 4-bit quantized LLama-2-13B-chat model.
We configure source containers to compiler and library versions close to those available on the system.
On Aurora, we have to manually patch the source code to compile with the Intel icpx compiler, as the SYCL backend cannot be compiled with Clang 21.
For XaaS source containers, we switch to the official oneAPI image at deployment.
In this benchmark, the specialized build performs comparably to \toolname{}, while the naive default build does not enable GPU (Figure~\ref{fig:llamacpp-execution time}).

\begin{figure}[t]
	\centering
	\includegraphics[width=\linewidth]{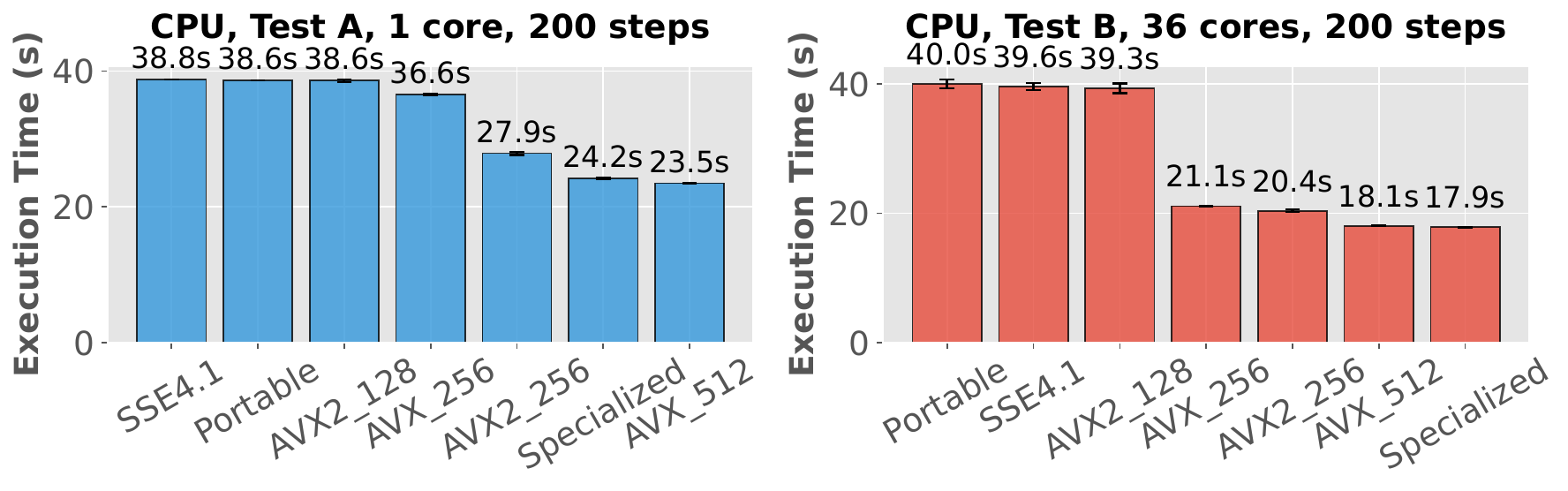}
	\includegraphics[width=\linewidth]{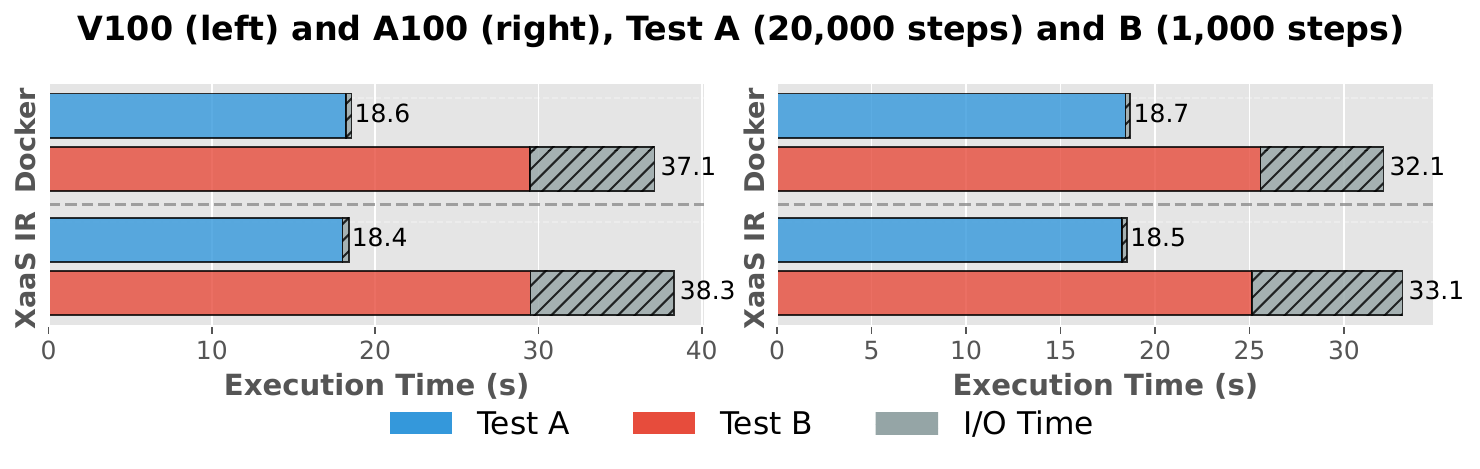}
	\caption{IR containers on CPU (top, Ault01-04) and GPU (bottom).}
	\vspace{-1.5em}
	\label{fig:gromacs-ir-container}
\end{figure}

\subsection{IR Containers}
We evaluate the performance portability of IR containers with GROMACS on CPU, with OpenMP, and tuned against five different CPU architectures, from SSE4.1 to AVX-512.
This requires discovering CPU tuning flags and delaying optimizations until final deployment (Section~\ref{sec:xaas-ir-containers-pipeline}).
For each variant, we build a separate container layer hosting \texttt{fftw3} library tuned for that architecture.
Additionally, we build IR containers with CUDA 12.8 by generating IR files with embedded device code, which we later lower to the target platform.
We compare against portable and specialized containers built with Clang 19 for SSE4.1 and AVX\_512 (CPU), and CUDA containers (GPU).
Figure~\ref{fig:gromacs-ir-container} shows the runtime of GROMACS with I/O time excluded, demonstrating that a specialization of the IR container can improve performance by up to 2x when compared to a performance-oblivious container.
We also evaluate a separate deployment of IR containers configured against CUDA and two vectorization levels for Ault23 and Ault25 nodes.
On the GPU, we provide performance comparable to a specialized container, with a slight increase in I/O time for test case B.

\emph{Configurability and System Dependency}
This experiment validates positively Hypotheses 1 and 2 on GROMACS (Section~\ref{sec:xaas-ir-containers}).
To deploy five GROMACS containers tuned to different ISAs, developers must build 8710 translation units (TU).
In our container, we build only 2695 IRs (69\% reduction).
The reduction would not be possible without the \toolname{} pipeline, as 96\% of compiled source files have incompatible build flags across projects; the primary reason is the inclusion of header files in the build directory.
Preprocessing determines that only 14.3\% of additional TUs require separate IR compilation.
However, 95\% of identical targets have different CPU tuning, which is resolved by the vectorization pass of our pipeline.

Four build configurations with two vectorization settings and CUDA require 7052 TUs, which we reduce to 2694 IRs (76\% reduction).
The build combinations obtained with enabling OpenMP and/or MPI require compiling and lowering 6976 TUs.
Thanks to preprocessing and determining when the OpenMP flag has no effect, we build only 2333 IRs (66.4\% reduction).

\vspace{-0.75em}

\subsection{Network Performance}
\label{sec:network_performance}
We focused on single-node deployments to demonstrate specialization to the available hardware and software,
and did not evaluate distributed execution due to technical limitations on our systems.
On Aurora, Apptainer containers did not function with MPI, and we had to use Thread-MPI instead.
On Clariden, co-location of MPI ranks is needed to utilize the four GH200 chips on each node.
However, the intra-node MPI communication is implemented separately from \emph{cxi}, the Slingshot provider in Libfabric~\cite{https://doi.org/10.1002/cpe.8203}.
Thus, containerized MPIs can access the high-speed network through Libfabric replacement, but cannot use shared memory.
While bare-metal Cray-MPICH achieves up to 64 GB/s on the same socket, co-located containers reach only up to 23.5 GB/s (OpenMPI).
LinkX~\cite{pritchard2023open} is a Libfabric provider that combines remote and local communication, and provides up to 64 (MPICH) and 70 (OpenMPI) GB/s of intra-node bandwidth.
However, the provider is still experimental, e.g., it does not function well on all benchmarks and hardware.

\section{Related Work}

\textbf{Building:}
%
% SC cut
%
Languages common in HPC, like C++ and Fortran, are notably missing in commonly used package managers.
EasyBuild~\cite{6495863} builds HPC applications from source using specific toolchains, supporting hierarchical module creation~\cite{7081225}.
Spack~\cite{10.1145/2807591.2807623} is a package manager that parameterizes builds with constraints and versioned dependencies.
Resolving dependencies can be reduced with declarative programming~\cite{10046107} or machine learning~\cite{10793143}.
E4S provides curated HPC software stacks, including hardware-specific containers~\cite{10513439,e4s}.
Binary distribution is possible: Spack uses binary caches~\cite{spackbinary}, and EESSI distributes EasyBuild stacks via network filesystems~\cite{https://doi.org/10.1002/spe.3075}.
\toolname{} complements these tools by addressing the trade-off between container portability and performance.

\textbf{Portable and HPC Containers:}
Injecting or replacing container libraries with host counterparts can be achieved with many container runtimes, but it can require expert knowledge of the system.
Apptainer~\cite{10.1371/journal.pone.0177459} supports semi-manual mounting of host MPI~\cite{apptainerdocs}.
Charliecloud~\cite{10.1145/3126908.3126925} uses heuristics to copy resource-specific files (NVIDIA, libfabric) into images, permanently modifying them.
Sarus~\cite{10.1007/978-3-030-34356-9_5, 10029965} and Podman-HPC~\cite{10030014} use OCI hooks to inject host MPI and GPU libraries.
\toolname{} can use the same hooks, but source containers can be compiled to use the version of the specialized library compatible with the one available on the system.
Vendor container registries offer optimized but platform-specific images~\cite{nvidiangc,amdinfinity}.

Containers can already contain intermediate representation as Python and Java bytecode~\cite{9242268}.
Popcorn Linux~\cite{10.1145/3093337.3037738} enables cross-ISA live migration with a custom compiler and kernel that transform LLVM IR into multi-ISA binaries with compatible data layouts~\cite{10.1145/3688351.3689152}.
H-Containers~\cite{10.1145/3381052.3381321, 10.1145/3524452} achieve migration by decompiling to LLVM IR and recompiling to different ISAs.
To the best of our knowledge, this is the only known use of IR for container deployment.
However, it differs fundamentally from \toolname{}: we use IR-based representations to access customized performance features of each system. 

\textbf{Performance Portability:}
Performance portability often involves rewriting applications using models like OpenMP, OpenACC, or SYCL~\cite{9484790, 10.1145/3529538.3529980}.
Frameworks provide new abstractions for memory access (Kokkos~\cite{CARTEREDWARDS20143202}), loop parallelism (Raja~\cite{8945721}), and data-centric programming (DaCe~\cite{10.1145/3458817.3476176}).
Compilers can translate programming idioms to specialized libraries~\cite{10.1145/3173162.3173182} and accelerators~\cite{10.1145/3578360.3580262},
and upgrade applications to use newer and specialized implementations of linear algebra libraries~\cite{8891611}.
\toolname{} focuses on portable representations of existing applications without rewriting or requiring single-source code.
We do not require applications to be single-source
or use the same set of source files across all systems and devices.

\textbf{Emulation, Translation, and JIT:}
Cross-ISA emulation, such as Docker with QEMU~\cite{dockerqemu}, is unsuitable for HPC due to performance overheads.
Runtime MPI ABI translation layers like\sloppy~Wi4MPI~\cite{9556086} can incur performance overhead.
Other tools include mpixlate~\cite{mpixlatecray} (compatibility with Cray MPI), MPITrampoline~\cite{schnetter_2022_6174409}, Mukautuva~\cite{mukautuva}, and MPI-Adapter2~\cite{10.1145/3636480.3637219}.
JIT compilation, used in CUDA PTX, OpenCL, SYCL IR~\cite{alpayOnePassBind2023}, allows for specialization of the final implementation by compiling the code dynamically.

\section{Discussion}

We demonstrate XaaS containers with representative HPC applications.
However, modern HPC workloads are often not limited to a single application~\cite{9309042}. 
Large HPC workflows like MOFA~\cite{yan2025mofadiscoveringmaterialscarbon} are built from several different tasks, each with its own requirements for CPU and GPU computation.
Performance-portable containers could create a seamless deployment of a heterogeneous workflow across different HPC hardware.
To transition the IR format to large and complex applications, we need to support dependency management (Section~\ref{sec:xaas-dependency}) and software installation (Section~\ref{sec:xaas-installation}).

\vspace{-1em}

\subsection{Dependency Management}
\label{sec:xaas-dependency}
When building different versions of an application, we should not repeat the entire build step for all dependencies.
Instead, dependencies should be composed into a final package, as is already the case for package managers such as Spack~\cite{10.1145/2807591.2807623}.
Standard containers distribute dependencies as binary images assembled for the selected specialization.
In IR containers, each dependency should be distributed in the IR form.
However, they must be located during the configuration phase of IR containers, and build parameters are affected by compilation and linking flags of dependencies.
This leads to a conflict: we want to deploy partially compiled applications, but still provide installation configuration to compose IR containers. %the generation of build configurations.

To support composability, future work can include creating \emph{fat} binaries with embedded IR, similarly to existing approaches for deploying GPU device code in CUDA and SYCL~\cite{cudacompilationtrajectory}.
This approach will generate a full installation target, allowing for seamless operation of build systems, while providing the necessary metadata and IRs to optimize and regenerate the dependency for the target.
Furthermore, extended dependency management could support version constraints, similarly to existing solutions in package managers.
This feature will restrict the matching process of specialization points, and prevent build failures caused by incompatibilities between the containerized application and its dependencies.
\vspace{-1em}

\subsection{Installation}
\label{sec:xaas-installation}
IR containers include the application's source code, even if the entire application is compiled to an intermediate representation.
This is not a strict requirement of our method but a limitation of existing build systems.
To finalize the application build, we need to perform linking and installation.
However, these steps can include many user-defined and customized instructions, such as generating custom headers, and they cannot be easily extracted from the build configuration.
Consequently, the IR image embeds all build configurations. % or recreates the selected one during deployment.
Automating installation would reduce the complexity of the container and solve the problem of IR container composability.
Furthermore, we can deduplicate installation targets as currently done with IR files.
Then, the IR container will only need to contain a shared installation core and a delta for each build configuration.

\section{Conclusions}

\toolname{} Source and Intermediate Representation (IR) containers bring a new methodology for software management in HPC.
We show that changing the software distribution allows for delaying performance critical decisions until the deployment, avoiding performance limitations of traditional containers.
Our prototype demonstrates that software deployments based on LLVM IR can significantly reduce the number of files that need to be generated without sacrificing performance.

\begin{acks}
	This project received support by the SwissTwins project (funded by the Swiss State Secretariat for Education, Research and Innovation) and the ERC PSAP project (Grant Agreement No. 101002047).
	This work was performed under the auspices of the U.S. Department of Energy by Lawrence Livermore National Laboratory under contract DE-AC52-07NA27344 (LLNL-CONF-2010482), Lawrence Livermore National Security, LLC, and by Argonne National Laboratory under
	Contract DE-AC02-06CH11357.
	We thank the Swiss National Supercomputing Centre (CSCS) and Argonne Leadership Computing Facility (ALCF) for providing computational resources and technical support that facilitated this project.
	The authors leveraged Claude to assist with light editing of the manuscript.
	Copilot and Claude Code were used during code development.
	All content and ideas remain the authors' original work.
\end{acks}

\bibliographystyle{ACM-Reference-Format}
\bibliography{serverless,own,hpc}

\appendix
\section{LLM Prompt to Discover Specialization Points}
\label{sec:llm-prompt}

I will share a build file, and I would like you to identify all the specialization points for an HPC program and the associated build flags used to enable those features during the build process. Please pay close attention to:

\begin{itemize}
    \item Comments and messages within the build file, as they often reveal the necessary flags.
    \item Functions like \texttt{gmx\_option\_multichoice}, which specify build flags and options for libraries.
    \item Ensure libraries are correctly matched to their corresponding build flags based on these functions.
    \item Option Commands: In some projects, build flags are provided in \texttt{option} commands. Look at these commands to extract the build flags correctly.
    \item Full Build Flags Extraction: Ensure that the full build flags are extracted, not just partial representations. For instance, if a flag is defined as \texttt{-DQE\_ENABLE\_CUDA=ON}, extract the entire flag with its value.
    \item Distinguish Between Build Flags and Preprocessor Definitions: Do not confuse preprocessor definitions (e.g., \texttt{\_\_CUDA}, \texttt{\_\_MPI}) with actual build flags (e.g., \texttt{-DQE\_ENABLE\_CUDA}, \texttt{-DQE\_ENABLE\_MPI}). Extract only the build flags that are explicitly defined in the build configuration.
    \item Portability Frameworks: Some build systems use portability frameworks like Kokkos. Pay attention to build flags like \texttt{-DKokkos\_ENABLE\_OPENMP}, \texttt{-DKokkos\_ENABLE\_PTHREAD}, and \texttt{-DKokkos\_ENABLE\_CUDA}.
    \item Vectorization Libraries: Some projects use external vectorization libraries like V4. Look for build flags such as \texttt{-DUSE\_V4\_ALTIVEC}, \texttt{-DUSE\_V4\_PORTABLE}, and \texttt{-DUSE\_V4\_SSE}.
\end{itemize}

Key Instructions:

1. Analyze Functions for Build Flags:

\begin{itemize}
    \item Look for functions such as \texttt{gmx\_option\_multichoice}, \newline \texttt{gmx\_dependent\_option}, and \texttt{option} commands that define build flags and their corresponding options.
    \item For example, the flag \texttt{-DGMX\_FFT\_LIBRARY} has options like \texttt{fftw3}, \texttt{mkl}, and \texttt{fftpack[built-in]}.
    \item Another example is \texttt{-DGMX\_GPU\_FFT\_LIBRARY} with options like \texttt{cuFFT}, \texttt{VkFFT}, \texttt{clFFT}, \texttt{rocFFT}, and \texttt{MKL}. Match the library names with the build flags from these function calls.
    \item Additionally, the flag \texttt{-DGMX\_GPU} has options like \texttt{CUDA}, \texttt{OpenCL}, \texttt{SYCL}, and \texttt{HIP}. Ensure these GPU backends are matched correctly to their corresponding flags.
    \item For Kokkos, look for flags like \texttt{-DKokkos\_ENABLE\_OPENMP}, \texttt{-DKokkos\_ENABLE\_PTHREAD}, and \texttt{-DKokkos\_ENABLE\_CUDA}.
\end{itemize}

2. Match Libraries to Flags:

\begin{itemize}
    \item Libraries should be matched to their respective build flags based on these function definitions.
    \item For example:
    \begin{itemize}
        \item If \texttt{GMX\_FFT\_LIBRARY} is set to \texttt{fftw3}, the build flag is \texttt{-DGMX\_FFT\_LIBRARY=fftw3}.
        \item If \texttt{GMX\_GPU\_FFT\_LIBRARY} is set to \texttt{cuFFT}, the build flag is \texttt{-DGMX\_GPU\_FFT\_LIBRARY=cuFFT}.
        \item For vectorization, look for flags like \texttt{-DUSE\_V4\_ALTIVEC}, \texttt{-DUSE\_V4\_PORTABLE}, and \texttt{-DUSE\_V4\_SSE}.
    \end{itemize}
\end{itemize}

3. Match GPU Backends to \texttt{GMX\_GPU}:

\begin{itemize}
    \item Ensure that GPU backends (CUDA, OpenCL, SYCL, HIP, METAL) are matched to the \texttt{GMX\_GPU} flag based on the \texttt{gmx\_option\_multichoice} function.
    \item For example:
    \begin{itemize}
        \item If \texttt{GMX\_GPU} is set to \texttt{CUDA}, the build flag is \texttt{-DGMX\_GPU=CUDA}.
        \item If \texttt{GMX\_GPU} is set to \texttt{SYCL}, the build flag is \texttt{-DGMX\_GPU=SYCL}.
    \end{itemize}
    \item For Quantum ESPRESSO: Ensure that GPU backends like CUDA are matched to their corresponding build flags, such as \texttt{-DQE\_ENABLE\_CUDA}, instead of preprocessor definitions like \texttt{\_\_CUDA}.
\end{itemize}

4. Consider Default Values and Dependencies:

\begin{itemize}
    \item Identify the default libraries and how they are conditionally set. For example:
    \begin{itemize}
        \item \texttt{GMX\_FFT\_LIBRARY\_DEFAULT} is \texttt{mkl} if \texttt{GMX\_INTEL\_LLVM} is set, otherwise \texttt{fftw3}.
        \item The GPU FFT library defaults vary based on the GPU backend (e.g., \texttt{cuFFT} for CUDA, \texttt{VkFFT} for OpenCL).
    \end{itemize}
\end{itemize}

5. Special Attention to FFT Libraries:

\begin{itemize}
    \item Look for all flags related to FFT libraries like:
    \begin{itemize}
        \item \texttt{-DGMX\_FFT\_LIBRARY}
        \item \texttt{-DGMX\_FFT\_LIBRARY\_DEFAULT}
        \item \texttt{-DGMX\_GPU\_FFT\_LIBRARY}
    \end{itemize}
    \item Extract not only the flag but also the corresponding library it enables (e.g., \texttt{fftw3}, \texttt{mkl}, \texttt{cuFFT}).
\end{itemize}

6. Include Relevant Build Flags:

\begin{itemize}
    \item Do not include preprocessor definitions generated internally. Only include build flags explicitly defined in the file.
    \item Ensure that each build flag is extracted with its full definition, including any assigned values.
\end{itemize}

Specifically, identify the following:

\begin{itemize}
    \item Does the build system support GPU builds? (For example, the presence of a flag like \texttt{BUILD\_GPU} indicates GPU support.)
    \item What GPU backends does it support (e.g. CUDA, HIP, SYCL, OpenCL)? Are these backends enabled or disabled by default? What is their minimum version, if specified?
    \item What parallel programming libraries (e.g. MPI, OpenMP, Pthread, thread-MPI, OpenACC) are supported, and are they enabled or disabled by default? What is their minimum version, if specified?
    \item What linear algebra libraries (e.g. BLAS, LAPACK, SCALAPACK, MKL/oneMKL) does the build system use, and under which conditions? What are the default libraries used in the build process?
    \item What Fast Fourier Transform libraries (e.g. FFTW, fftpack, MKL/oneMKL, cuFFT, vkFFT, clFFT, rocFFT) does the build system use? What library is built-in? Are there specific dependencies for the library to be used (for example, they must be used with a certain GPU backend or parallel library)? Are they enabled or disabled by default? For the build-flags, look for flags defined via \texttt{gmx\_option\_multichoice} such as \texttt{-DGMX\_FFT\_LIBRARY}, \texttt{-DGMX\_FFT\_LIBRARY\_DEFAULT}, 
    \newline \texttt{-DGMX\_GPU\_FFT\_LIBRARY}.
    \item What other external libraries are used, what versions are specified, and what are their dependencies? List all external libraries and the conditions for their use.
    \item What other compiler flags are supported?
    \item Are there build flags used to optimize the performance of the program? (e.g., auto-tuning, team reduction, hierarchical parallelism, accumulators, qunatization, batch size, force use of custom matrix multiplications)
    \item Which compilers are supported, and what are the minimum versions required?
    \item What architectures does the system support?
    \item Does it support SIMD vectorization, and what vectorization levels are supported? find the build flag for each supported vectorization level.
    \item What is the minimum version required for the build system? Is it a CMake or Make build system?
    \item Are there any libraries that require internal builds? If so, name them and provide the build flags (e.g. \texttt{-DGMX\_BUILD\_OWN\newline\_FFTW}, \texttt{DBUILD\_INTERNAL\_KOKKOS}).
\end{itemize}

The answer should be provided as a JSON structure adhering to the specified schema, with keys including \texttt{gpu\_build}, \texttt{gpu\_backends}, \texttt{parallel\_programming\_libraries}, \texttt{linear\_algebra\_libraries}, \texttt{fft\_libraries}, \texttt{other\_external\_libraries}, \texttt{optimization\_build\_flags}, \texttt{compiler\_flags}, \texttt{compilers}, \texttt{architectures}, \texttt{simd\_vectorization}, and \texttt{build\_system}, \texttt{internal\_build}. The \texttt{build\_flag} value for each feature should be the flag itself (e.g., \texttt{-DGMX\_SIMD}, \texttt{-DGMX\_GPU}, \texttt{-DQE\_ENABLE\_CUDA}, \texttt{-DQE\_ENABLE\_MPI}, \texttt{-DKokkos\_ENABLE\_OPENMP}, \texttt{-DUSE\_V4\_ALTIVEC}) without any surrounding text. Do not include any preprocessor definitions that are generated internally. The response must be a valid JSON structure; do not include any introductory or explanatory text.

Here is the build file: \texttt{\{file\_content\}}

JSON output schema. Use this JSON schema to format your response but do not include it in the output: \texttt{\{schema\}}
\section{JSON Schema for Specialization Points}
\label{sec:json-schema}

\lstdefinestyle{jsonclean}{
  language=json,
  breaklines=true,
  breakindent=0pt,
  basicstyle=\ttfamily\scriptsize,
  numbers=none,
  frame=none,
  keywordstyle=\color{black},
  stringstyle=\color{black},
  commentstyle=\color{black},
  identifierstyle=\color{black}
}

\begin{lstlisting}[style=jsonclean]
{
    "$schema": "http://json-schema.org/draft-07/schema#",
    "type": "object",
    "properties": {
      "gpu_build": {
        "type": "object",
        "properties": {
          "value": {
            "type": "boolean"
          },
          "build_flag": {
            "type": ["string", "null"]
          }
        },
        "required": ["value", "build_flag"]
      },
      "gpu_backends": {
        "type": "object",
        "additionalProperties": {
          "type": "object",
          "properties": {
            "used_as_default": {
              "type": "boolean"
            },
            "build_flag": {
              "type": ["string", "null"]
            }, 
            "minimum_version": {
              "type": ["string", "null"]
            }
          },
          "required": ["used_as_default", "build_flag", "minimum_version"]
        }
      },
      "parallel_programming_libraries": {
        "type": "object",
        "additionalProperties": {
          "type": "object",
          "properties": {
            "used_as_default": {
              "type": "boolean"
            },
            "build_flag": {
              "type": ["string", "null"]
            }, 
            "minimum_version": {
              "type": ["string", "null"]
            }
          },
          "required": ["used_as_default", "build_flag", "minimum_version"]
        }
      },
      "linear_algebra_libraries": {
        "type": "object",
        "additionalProperties": {
          "type": "object",
          "properties": {
            "used_as_default": {
              "type": "boolean"
            },
            "build_flag": {
              "type": ["string", "null"]
            }, 
            "condition": {
              "type": ["string", "null"]
            }
          },
          "required": ["used_as_default", "build_flag", "condition"]
        }
      },
      "FFT_libraries": {
        "type": "object",
        "additionalProperties": {
          "type": "object",
          "properties": {
            "built-in": {
              "type": "boolean"
            },
            "used_as_default": {
              "type": "boolean"
            },
            "dependencies": {
              "type": ["string", "null"]
            },
            "build_flag": {
              "type": ["string", "null"]
            }
          },
          "required": ["used_as_default", "condition", "build_flag"]
        }
      },
      "other_external_libraries": {
        "type": "object",
        "additionalProperties": {
          "type": "object",
          "properties": {
            "version": {
              "type": "string"
            },
            "used_as_default": {
              "type": "boolean"
            },
            "conditions": {
              "type": "string"
            },
            "build_flag": {
              "type": ["string", "null"]
            }
          },
          "required": ["version", "used_as_default", "conditions", "build_flag"]
        }
      },
      "compiler_flags": {
        "type": "array",
        "items": {
          "type": "string"
        }
      },
      "optimization_build_flags": {
        "type": "array",
        "items": {
          "type": "string"
        }
      },
      "compilers": {
        "type": "object",
        "additionalProperties": {
          "type": "object",
          "properties": {
            "minimum_version": {
              "type": "string"
            }
          },
          "required": ["minimum_version"]
        }
      },
      "architectures": {
        "type": "array",
        "items": {
          "type": "string"
        }
      },
      "simd_vectorization": {
        "type": "object",
        "additionalProperties": {
          "type": "object",
          "properties": {
            "build_flag": {
              "type": ["string", "null"]
            },
            "default": {
              "type": "boolean"
            }
          },
          "required": ["build_flag", "default"]
        }
      }, 
        "build_system": {
          "type": "object",
          "properties": {
            "type": {
              "type": "string",
              "enum": ["cmake", "make", "undetermined"]
            },
            "minimum_version": {
              "type": "string"
            }
          },
          "required": ["type", "minimum_version"]
        }, 
        "internal_build": {
        "type": "object",
        "properties": {
          "library_name": {
            "type": "string"
          },
          "build_flag": {
            "type": ["string", "null"]
          }
        },
        "required": ["library_name", "build_flag"]
    }
    },
    "required": [
      "gpu_build",
      "gpu_backends",
      "parallel_programming_libraries",
      "linear_algebra_libraries",
      "FFT_libraries",
      "other_external_libraries",
      "compiler_flags",
      "optimization_build_flags",
      "compilers",
      "architectures",
      "simd_vectorization", 
       "build_system",
       "internal_build"
    ],
    "additionalProperties": false
}
\end{lstlisting}

\end{document}